\documentclass[twocolumn]{pasj00}

\begin{document}
\SetRunningHead{Y. Takeda et al.}{Beryllium Abundances of Solar-Analogs}
\Received{2010/10/20}
\Accepted{2011/03/22}

\title{Beryllium Abundances of Solar-Analog Stars
\thanks{Based on data collected at Subaru Telescope, which is operated
by the National Astronomical Observatory of Japan. }
\thanks{The large data are separately given in the electronic form 
(tables E1, E2, E3), which will be available at the E-PASJ web site 
upon publication.}
}

%

\author{
Yoichi \textsc{Takeda},\altaffilmark{1}
Akito \textsc{Tajitsu},\altaffilmark{2}
Satoshi \textsc{Honda},\altaffilmark{3}
Satoshi \textsc{Kawanomoto},\altaffilmark{1}\\
Hiroyasu \textsc{Ando},\altaffilmark{1} and
Takashi \textsc{Sakurai}\altaffilmark{1}}
\altaffiltext{1}{National Astronomical Observatory, 2-21-1 Osawa, Mitaka, Tokyo 181-8588}
\email{takeda.yoichi@nao.ac.jp, kawanomoto.satoshi@nao.ac.jp, ando.hys@nao.ac.jp,\\ 
sakurai@solar.mtk.nao.ac.jp}
\altaffiltext{2}{Subaru Telescope, 650 North A'ohoku Place, Hilo, Hawaii 96720, U.S.A.}
\email{tajitsu@subaru.naoj.org}
\altaffiltext{3}{Kwasan Observatory, Graduate School of Science, Kyoto University,\\
17 Ohmine-cho Kita Kazan, Yamashina-ku, Kyoto 607-8471}
\email{honda@kwasan.kyoto-u.ac.jp}

\maketitle

\begin{abstract}

An extensive beryllium abundance analysis was conducted 
for 118 solar analogs (along with 87 FGK standard stars) 
by applying the spectrum synthesis technique to the near-UV region 
comprising the Be~{\sc ii} line at 3131.066~$\rm\AA$, in an attempt 
to investigate whether Be suffers any depletion such as the case 
of Li showing a large diversity. We found that, 
while most of these Sun-like stars are superficially similar 
in terms of their $A$(Be) (Be abundances) around the solar value 
within $\sim \pm 0.2$~dex, 4 out of 118 samples turned out 
strikingly Be-deficient (by more than 
$\sim 2$~dex) and these 4 stars belong to the group of lowest 
$v_{\rm e}\sin i$ (projected rotation velocity). 
Moreover, even for the other majority showing an apparent similarity 
in Be, we can recognize a tendency that $A$(Be) gradually increases 
with an increase in $v_{\rm e}\sin i$.
These observational facts suggest that any solar analog star
(including the Sun) generally suffers some kind of Be depletion
during their lives, where the rotational velocity (or the angular 
momentum) plays an important role in the sense that depletion
tends to be enhanced by slower rotation. Hence, our findings require
that the occasionally stated view ``G-type dwarfs with 
$T_{\rm eff} \ltsim 6000$~K are essentially homogeneous in Be 
with their original composition retained'' should be revised. 
Also, our analysis indicates that the difference of $\sim 0.2$~dex 
in $A$(Be) between the solar photosphere and the meteorite
really exists, implying that ``UV missing opacity'' is irrelevant
at least for this Be~{\sc ii} line.

\end{abstract}

%


\section{Introduction}

This paper is the third of the series, resulting from our recent 
research project on a large homogeneous sample of solar analog stars, 
which aims to clarify the behavior of their Li abundances in connection 
with the stellar properties such as rotation, stellar activity, and age.
We now here focus on beryllium, another light element similar to lithium.

\subsection{Lithium in Solar-Analog Stars}

As Li nuclei are burned and destroyed when they are conveyed 
into the hot stellar interior ($T \gtsim 2.5 \times 10^{6}$~K),
we can gain from its surface composition valuable information regarding 
the past history and the physical mechanism of stellar envelope mixing. 
Especially, the behavior of $A$(Li)\footnote{Throughout this study, 
$A$(X) means the logarithmic (number) abundance of any element (X) in 
the usual normalization of $A$(H) = 12, whereas we express
the logarithmic differential abundance of X relative to the Sun by 
[X/H] ($\equiv A_{\rm star}$(X) $-$ $A_{\odot}$(X)).} (Li abundances) in 
Sun-like stars is of particular interest, as it poses intriguing puzzles: \\
---(1) Why $A$(Li)'s show a remarkably large diversity ($\sim 2$~dex) 
even for stars with very similar parameters (e.g., Takeda \& Kawanomoto 
2005)? \\
---(2) Why planet-host stars show markedly lower $A$(Li) than 
non-planet-host stars around the solar $T_{\rm eff}$ region 
(e.g., Israelian et al. 2009)\footnote{Critical arguments against 
the existence of such a characteristic behavior of Li specific to
planet-host stars also exist (e.g., Baumann et al. 2010).}? \\
Since such characteristic trends are difficult to explain by the naive 
classical picture of surface Li being determined by age (duration 
time of gradual Li depletion by way of convective mixing) and 
$T_{\rm eff}$ (affecting the depth of convection zone), it has been 
the central task to find the key parameter(s) responsible for 
these observed facts.

In an attempt toward clarifying this issue, Takeda et al. (2007; 
hereinafter referred to as Paper I) conducted an extensive study on 
the connection of Li abundances and stellar parameters for 118 solar 
analogs and found that $A$(Li) was found to closely correlate with 
the line-width (combination of rotation and macroturbulence), from
which they suspected that rotation may be the key factor.  
Successively, Takeda et al. (2010; hereinafter referred to as Paper II ) 
studied the residual flux at the core of the strong Ca~{\sc ii} 8542 
line ($r_{0}$(8542); a good indicator of stellar activity,
considered to be rotation-dependent because of its dynamo origin) for 
these 118 stars and confirmed that $r_{0}$(8542) shows a positive connection 
with the line width as well as $A$(Li). Thus, there is no doubt that 
the stellar angular momentum plays the decisive role in controlling 
the surface Li abundances of Sun-like stars, in the sense that
Li tends to be more depleted for slowly rotating stars.
(See, e.g., Pinsonneault 2010 for a review regarding the impact of 
rotatinal mixing on the surface Li abundance of late-type stars.)

\subsection{Beryllium: Another Probe}

So, the next task should be to find a theoretical explanation
reasonably accounting for the observational fact.
Interestingly, Bouvier (2008) recently proposed a promising 
mechanism that slow rotation can produce an efficient mixing 
as well as Li depletion. He showed in his theoretical simulation
that slow rotators develop a high degree of differential rotation
between the radiative core and the convective envelope, which
eventually promotes lithium depletion by enhanced mixing,
while fast rotators evolve with little such core--envelope
decoupling. If such a picture is correct, the key to understanding
the mixing mechanism lies in ``the bottom of the convection zone,''
where shear instability induced by differential rotation
takes place in slow rotators. 

We would point out that the best way to probe the condition of 
this deep layer is to study the abundance of Be (destroyed at 
temperature of $T  \gtsim 3.5 \times 10^{6}$ K; i.e., somewhat 
higher than the Li-burning temperature) along with that of Li.
According to figure 1, where the $T$--$P$ and 
$F_{\rm conv}/F_{\rm tot}$--$P$ relations of the 
standard solar model (Stix 2002) are depicted, the critical
temperatures of Li as well as Be burning just lie near to
the base of the convection zone. Therefore, we would be able 
to extract useful information on this critical layer by 
examining/comparing the abundances of these two elements for a 
large sample of solar analogs having a wide variety of 
rotational velocities, which might serve as a
good touchstone for the theory of mixing mechanism.

\setcounter{figure}{0}
\begin{figure}
  \begin{center}
    \FigureFile(80mm,80mm){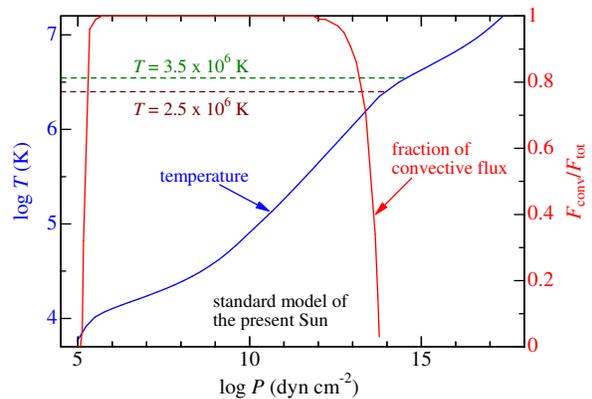}
  \end{center}
\caption{
Temperature ($T$) and fraction of convective flux
($F_{\rm conv}/F_{\rm total}$) for the standard solar model
(Stix 2002), plotted against pressure ($P$).
The critical temperatures of Li-burning ($T \sim 2.5 \times 10^{6}$~K) 
and Be-burning ($T \sim 3.5 \times 10^{6}$~K) are indicated by 
horizontal dashed lines.
}
\end{figure}

\subsection{Controversy on Be in Solar-Type Stars}

Unfortunately, any consensus has not yet been accomplished concerning
the behavior of Be in solar-analog stars.\\
--- Santos et al. (2004a) suggested from their extensive Be abundance 
analysis of FGK stars (6300~K~$\gtsim T_{\rm eff} \gtsim 4800$~K)
that a significant Be depletion in a narrow Teff range (similar to the 
Li gap for F stars) may exist in solar-temperature stars.\\ 
--- Meanwhile, Randich (2010) argued that 
Be abundances are generally uniform (i.e., with almost no sign 
of depletion) in G-type dwarfs with $T_{\rm eff} \ltsim 6000$~K 
regardless of the large spread of Li abundances (in contrast to 
F-type stars with $T_{\rm eff} \gtsim 6000$~K showing appreciable Be 
deficiencies in accordance with Li), which means that Li and Be 
do not have any correlation in Sun-like stars. \\
--- Takeda and Tajitsu (2009) determined the Li and Be abundances 
for three solar twins (HIP~56948, HIP~79672, and HIP~100963) along 
with the Sun, and concluded that all three have essentially the same 
$A$(Be) with the solar value, in spite of the marked difference 
in $A$(Li).\\
--- Stephens et al. (1997) reported, in their extensive Be abundance 
investigation for $\sim$~60 Li-deficient stars, that some early-G stars 
at $T_{\rm eff} \ltsim 6000$~K exhibit moderate Be deficiencies 
by a factor of $\sim$~2--4 (apart from the considerable Be depletion 
for F stars at 6000~K~$\ltsim T_{\rm eff} \ltsim$~6800~K).\\
--- Boesgaard and Hollek's (2009) Be abundance study for 50 
``one solar mass'' stars (6500~K~$\gtsim T_{\rm eff} \gtsim 5600$~K)
revealed several 3--4 very Be-depleted stars despite that most stars
are quite homogeneous in [Be/Fe]; however, since these outliers
are either subgiants or $T_{\rm eff} > 6000$~K dwarfs, it still 
remains an open question whether Be depletion occurs in solar analogs.

Hence, it is still unsettled in which condition Be is depleted (or not 
depleted) along with Li in G-dwarfs similar to the Sun.
While clarifying this issue is requisite for understanding the mechanism 
how the depletion takes place (e.g., the penetration depth of mixing 
at the bottom of the convection zone), the available observational data 
do not seem to be sufficient, because systematic Be abundance study 
specifically directed to a large homogeneous sample of ``solar analog 
stars'' has never been carried out to our knowledge.

\subsection{Purpose of This Study}

Given this situation, we decided to determine the Be abundances of 
118 solar analogs, for which Li abundances and various stellar 
parameters have already been well established in Papers I and II, 
by using the Be~{\sc ii} doublet lines at 3130--3131~$\rm\AA$
based on the high-dispersion near-UV spectra obtained by
Subaru/HDS (just suitable because of its sufficiently high
efficiency down to $\lambda \sim 3000 \rm\AA$ thanks to the
near-UV-sensitive CCD as well as the high-transparency at
Mauna Kea summit), in order to investigate whether any meaningful 
correlation exists between $A$(Be) and $A$(Li), and whether
they show any dependence upon the stellar physical parameters 
(e.g., stellar rotation, $T_{\rm eff}$, $age$, activity index, etc.).  
This is the aim of this investigation.

The remainder of this paper is organized as follows.
After describing the targets and the observational data (section 2), 
we explain the details of abundance determinations in section 3, 
followed by section 4 where the results of Be abundances are 
examined and discussed, first for FGK standard stars and then 
for solar analog stars. The conclusion is summarized in section 5.
In addition, an appendix is prepared where the non-LTE effect 
on Be abundance determination from the Be~{\sc ii} 3131 line 
is mentioned.

\section{Observational Data}

\subsection{Target Stars}

The main targets in this investigation are the 118 solar analogs
studied in detail in Papers I and II, for which the atmospheric parameters
[the effective temperature ($T_{\rm eff}$), surface gravity ($\log g$),
microturbulent velocity dispersion ($v_{\rm t}$), and metallicity 
(represented by Fe abundance relative to the Sun; [Fe/H])], the
lithium abundance ($A$(Li)), and the stellar parameters [the luminosity ($L$), 
mass ($M$), and age ($age$)] were already established (cf. sections 3 and 4 
in Paper I). Besides, the residual flux at the line center of the 
Ca~{\sc ii} 8542 ($r_{0}$(8542); good indicator of stellar activity) 
and the projected rotational velocity ($v_{\rm e}\sin i$) were also 
evaluated in Paper II (cf. sections 2 and 3 therein).

In addition, we also included 87 FGK stars 
(mostly dwarfs but including several subgiants) selected from 160 
standard stars,  which were studied in detail by Takeda et al. 
(2005; atmospheric parameters), Takeda and Kawanomoto (2005; 
Li abundances), and Takeda (2007; elemental abundances and stellar 
parameters such as $age$). These stars (ranging in 
7000~K~$\gtsim T_{\rm eff} \gtsim$~5000~K) were added mainly for 
the purpose of checking the behavior of Be in the wider range of 
$T_{\rm eff}$ by comparing it with the published results. 

Three kinds of representative parameter correlations 
($\log g$ vs. $T_{\rm eff}$, 
$\log L$ vs. $\log T_{\rm eff}$, and [Fe/H] vs. $\log age$)
for the program stars are depicted in figure 2, where we can recognize
a remarkable similarity of the parameters as well as a rather tight
age--metallicity relation for the 118 solar analogs (black
filled circles). The fundamental data (magnitudes, colors, stellar parameters,
etc.) for all our target stars are summarized in electronic table E1.

\setcounter{figure}{1}
\begin{figure}
  \begin{center}
    \FigureFile(70mm,120mm){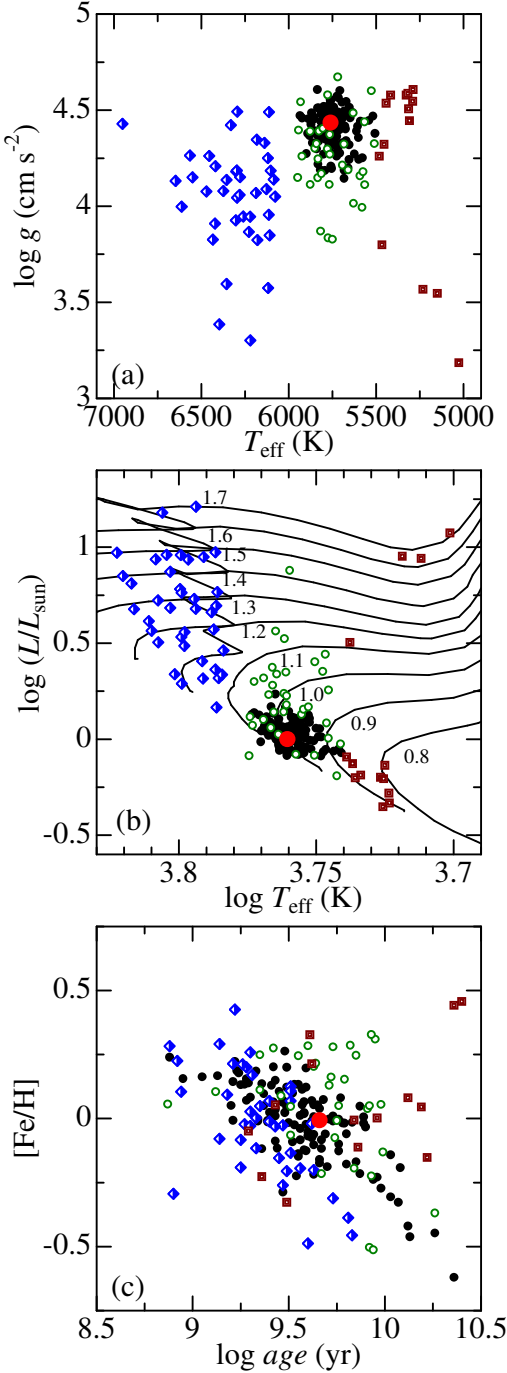}
  \end{center}
\caption{
Correlations of representative stellar parameters 
for the program stars: (a) Surface gravity vs. effective temperature,
(b) luminosity vs. effective temperature (Girardi et al.'s (2000)
theoretical evolutionary tracks for the solar metallicity stars of 
0.8~$M_{\odot}$ to 1.7~$M_{\odot}$ are also shown), and
(c) age vs. metallicity. Filled (black) circles $\cdots$ 118 solar 
analogs, half-filled (blue) diamonds $\cdots$ 39 standard stars
of F-type (7000~K~$> T_{\rm eff} > 6000$~K); open (green) circles
$\cdots$ 34 standard stars of early G type 
(6000~K~$> T_{\rm eff} > 5500$~K); outlined (brown) squares $\cdots$
14 standard stars of late G- or early K-type 
(5500~K~$> T_{\rm eff} > 5000$~K). The large (red) circle indicates 
the Sun.
}
\end{figure}

\subsection{Observations}

The observations of 205 targets (118 solar analogs and 87 standard
stars) and Vesta (substitute for the Sun)  were carried out on the 4
nights (Hawaii Standard Time) of 2009 August 6, 2009 November 27, 
2010 February 4, and 2010 May 24 with the High Dispersion Spectrograph 
(HDS; Noguchi et al. 2002) placed at the Nasmyth platform of the 
8.2-m Subaru Telescope, by which high-dispersion spectra 
covering $\sim$~3000--4600~$\rm\AA$ could be obtained with two CCDs 
of 2K$\times$4K pixels in the standard Ub setting with 
the blue cross disperser. 
The spectrum resolving power was $R \simeq 60000$ with the slit width 
set at $0.''6$ (300 $\mu$m) and a binning of 2$\times$2 pixels. 
The typical integrated exposure times were $\sim$~10--20~min for solar 
analogs (typically $V \sim 8$) and $\sim$~several min for standard stars
(typically $V \sim 5$). 
The basic data of these spectra (observing date, exposure time, 
S/N ratio, etc.) are given in electronic table E1.

\subsection{Data Reduction}

The reduction of the spectra (bias subtraction, flat-fielding, 
scattered-light subtraction, spectrum extraction, wavelength calibration,
co-adding of frames to improve S/N, continuum normalization) was 
performed by using the ``echelle'' package of 
the software IRAF\footnote{IRAF is distributed
    by the National Optical Astronomy Observatories,
    which is operated by the Association of Universities for Research
    in Astronomy, Inc. under cooperative agreement with
    the National Science Foundation.} 
in a standard manner. 
Typical S/N ratios of $\sim 100$ were attained at the position of 
Be~{\sc ii} 3130--3131 doublet lines in the finally resulting spectra 
for most of the targets.

\section{Beryllium Abundance Determination}

\subsection{Basic Strategy}

Since Be~{\sc ii} 3130--3131 doublet lines, on which we rely for
Be abundance determinations, are located in the UV region heavily crowded 
with a number of spectral lines, applying the spectrum synthesis 
technique is requisite. Regarding the data of atomic/molecular lines,
we exclusively adopted Primas et al.'s (1997) line list.
Figure 3a shows a comparison of theoretical and observed (flux) spectrum
of the Sun in the 3128--3133~$\rm\AA$ region, while the used line data 
are presented in table 1 along with the information of each line's strength.
(Note that, in figures 3a and b, the calculation was carried out
with the standard solar abundances and no attempt was made to fit the 
theoretical spectrum with the observed one by adjusting the abundances.)

Inspecting figure 3a and table 1, we decided to use only the 
Be~{\sc ii} 3131.066 line (weaker one of the doublet), since 
the stronger Be~{\sc ii} 3130.421 line is unsuitable because of being 
heavily blended with nearby V~{\sc ii} or OH lines. 
Here, the critical point we should attend is that Be~{\sc ii} 
3131.066 is blended with Fe~{\sc i} 3131.043,\footnote{
Although this line has not been confirmed from atomic physics 
but just empirically assumed by Primas et al. (1997) in order to 
reproduce the real spectrum of a Be-depleted star, we regard that this 
identification is surely reasonable (cf. subsection 3.3).
We should note, however, that other possibilities of line blending with 
the Be~{\sc ii}~3131.066 line have also been considered in previous studies. 
For example, Garc\'{\i}a L\'{o}pez et al. (1995b) included Mn~{\sc i} 
3131.037 line ($\chi_{\rm low} = 3.77$~eV) along with its appropriately 
adjusted $\log gf$, instead of the Fe~{\sc i} 3131.043 we adopted.
Further, King et al. (1997) carefully examined the blending effect of
other lines such as those of Mn~{\sc ii} or Ti~{\sc ii} or CH 
(in addition to this Mn~{\sc i} 3131.037 line) with an aim to reproduce 
the blue wing of the Be~{\sc ii}~3131.066 line.
} 
the contribution of the latter has to be removed for reliable 
Be abundance determination. 

Therefore, we focused on the 3130.65--3131.35~$\rm\AA$ region (cf. 
figure 3b) comprising three conspicuous features: the left one is due 
to blends of Nb~{\sc ii} 3130.780 and Ti~{\sc ii} 3130.810 lines, the 
middle one is due to Fe~{\sc i} 3131.043 and Be~{\sc ii} 3131.066, and 
the right one is primarily due to Fe~{\sc i} 3131.235 line (actually 
a composite of Cr~{\sc i}, Fe~{\sc i}, and Tm~{\sc ii} lines). 
We may then hope that the Be abundance can be reasonably established, 
since the contribution of Fe~{\sc i} 3131.043 may be controlled by 
simultaneously considering the Fe~{\sc i} 3131.235 line, while
the Nb~{\sc ii}+Ti~{\sc ii} feature contains information to regulate
the line broadening. Accordingly, our abundance determination
is based on matching between theoretical and observed 
spectra in this 3130.65--3131.35~$\rm\AA$ region by adjusting the 
abundances of these 4 elements (Be, Cr, Fe, Nb) along with the 
line broadening function.

\setcounter{figure}{2}
\begin{figure}
  \begin{center}
    \FigureFile(80mm,80mm){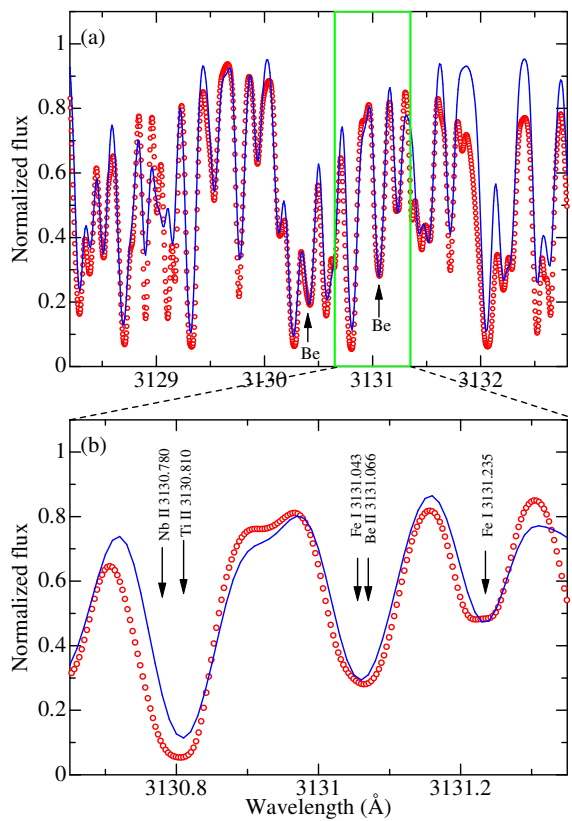}
  \end{center}
\caption{Theoretically synthesized spectrum for the Sun (blue solid line), 
where the $gf$ values in table 1 were used along with Anders and 
Grevesse's (1989) standard solar photospheric abundances (except for Fe, 
for which we adopted 7.50), compared with Kurucz et al.'s (1984) 
solar flux spectrum (red open circles). (a) 3128.2--3132.8~$\rm\AA$
region; (b) 3130.65--3131.35~$\rm\AA$ region (magnification of (a)).
}
\end{figure}

\subsection{Spectrum Fitting Analysis}

Practically, we used a stellar spectrum analysis tool ``MPFIT'', which 
was developed based on Kurucz's (1993) ATLAS9/WIDTH9 program and 
has a function of establishing the spectrum-related parameters 
(elemental abundances, macrobroadening parameters, radial velocity, 
etc.) by automatically searching for the best-fit solutions 
without any necessity of precisely placing the continuum level 
(Takeda 1995). 

We interpolated Kurucz's (1993) grid of ATLAS9 model atmospheres 
in terms of $T_{\rm eff}$, $\log g$, and [Fe/H] to generate the 
atmospheric model for each star. 
The assumption of LTE (local thermodynamical equilibrium) was adopted 
throughout this study, since the non-LTE effect on Be abundance 
determination from the Be~{\sc ii} resonance doublet
at 3030--3031~$\rm\AA$ is insignificant (almost negligible especially 
for solar-analog stars) as separately discussed in Appendix. 
Besides, the effect of atmospheric inhomogeneity (3D effect) is also 
unimportant for the deep-forming Be~{\sc II} lines according to 
Asplund (2005), by which the use of classical treatment is 
justified.

Thus, given the atmospheric model, we determined for each star
the abundances of four elements [$A$(Be), $A$(Ti), $A$(Fe), 
and $A$(Nb)],\footnote{The contributions of the lines of other 
elements than these four (cf. table 1) were formally 
included by assuming the solar abundances scaled with the metallicity; 
i.e., $A$(el) = $A_{\odot}$(el) + [Fe/H].}
along with the macrobroadening parameter 
($v_{\rm M}$; $e$-folding half-width of the Gaussian macrobroadening
function, $f_{\rm M}(v) \propto \exp [(-v/v_{\rm M})^{2}]$) and the
radial velocity shift, by applying the MPFIT program 
to the observed spectrum in the 3130.65--3131.35~$\rm\AA$ region. 
The finally established abundances of $A$(Be) and $A$(Fe) 
are given in electronic table E2, and the theoretical spectra 
corresponding to the best-fit solutions (along with the observed 
spectra) are shown in figures 4 (solar analogs) and 5 (FGK standards). 

\subsection{EW Evaluation}

Actually, there were cases where abundance solutions
did not converge. When any of $A$(Ti), $A$(Fe), or $A$(Nb)
was indeterminable, we fixed it at the metallicity-scaled solar
abundance [i.e., $A$(Ti) = $A_{\odot}$(Ti) + [Fe/H], etc.]
and retried the iteration procedure for the best-fit solutions.
Meanwhile, when $A$(Be) could not be determined (Be-depleted case
of too weak Be~{\sc ii} line), 
we neglected its contribution by assuming $A$(Be) = $-9.99$, and 
retried the iteration. In this case, the upper limit of equivalent 
width for the Be~{\sc ii} 3131.066 line ($EW_{\rm Be II\; 3131}^{\rm UL}$) 
was evaluated as $EW_{\rm Be II\; 3131}^{\rm UL} \simeq k \times$~FWHM/(S/N),
where $k$ is a factor we assumed to be 2 according to our 
experience, S/N is the signal-to-noise ratio ($\sim 100$), 
and FWHM was estimated from $v_{\rm M}$ as 
FWHM~$\simeq 2\sqrt{\ln 2} \; (\lambda v_{\rm M}/c)$ ($c$: velocity 
of light). We then derived the upper limit of $A$(Be) from 
such obtained $EW_{\rm Be II\; 3131}^{\rm UL}$.

Whether the contribution of the Fe~{\sc i} 3131.043 line could be 
successfully removed can be checked by comparing $A$(Fe) derived 
from this fitting and that already established in Paper I by using
a number of Fe lines. This comparison is displayed in figure 6a,
from which we can confirm that both are mostly consistent
to within $\ltsim 0.2$~dex.
We also computed $EW_{\rm Be II\; 3131}$ and  $EW_{\rm Fe I\; 3131}$
``inversely'' from $A$(Be) and $A$(Fe) (resulting from the spectrum 
synthesis analysis) along with the adopted atmospheric model/parameters, 
in order to examine to which extent the Fe~{\sc i} line
contributes to the (Fe~{\sc i} +) Be~{\sc ii} feature.
According to figure 6b, the strength of this Fe~{\sc i} line
is weaker than that of Be~{\sc ii} line in most cases (typically
$\sim$~30--40\% for solar analogs), though the former can outweigh
(or even becomes predominant over) the latter for the Be-depleted
cases particularly seen for F-type stars. These two kinds of $EW$ 
values are also given in electronic table E2.

\setcounter{figure}{5}
\begin{figure}
  \begin{center}
    \FigureFile(70mm,120mm){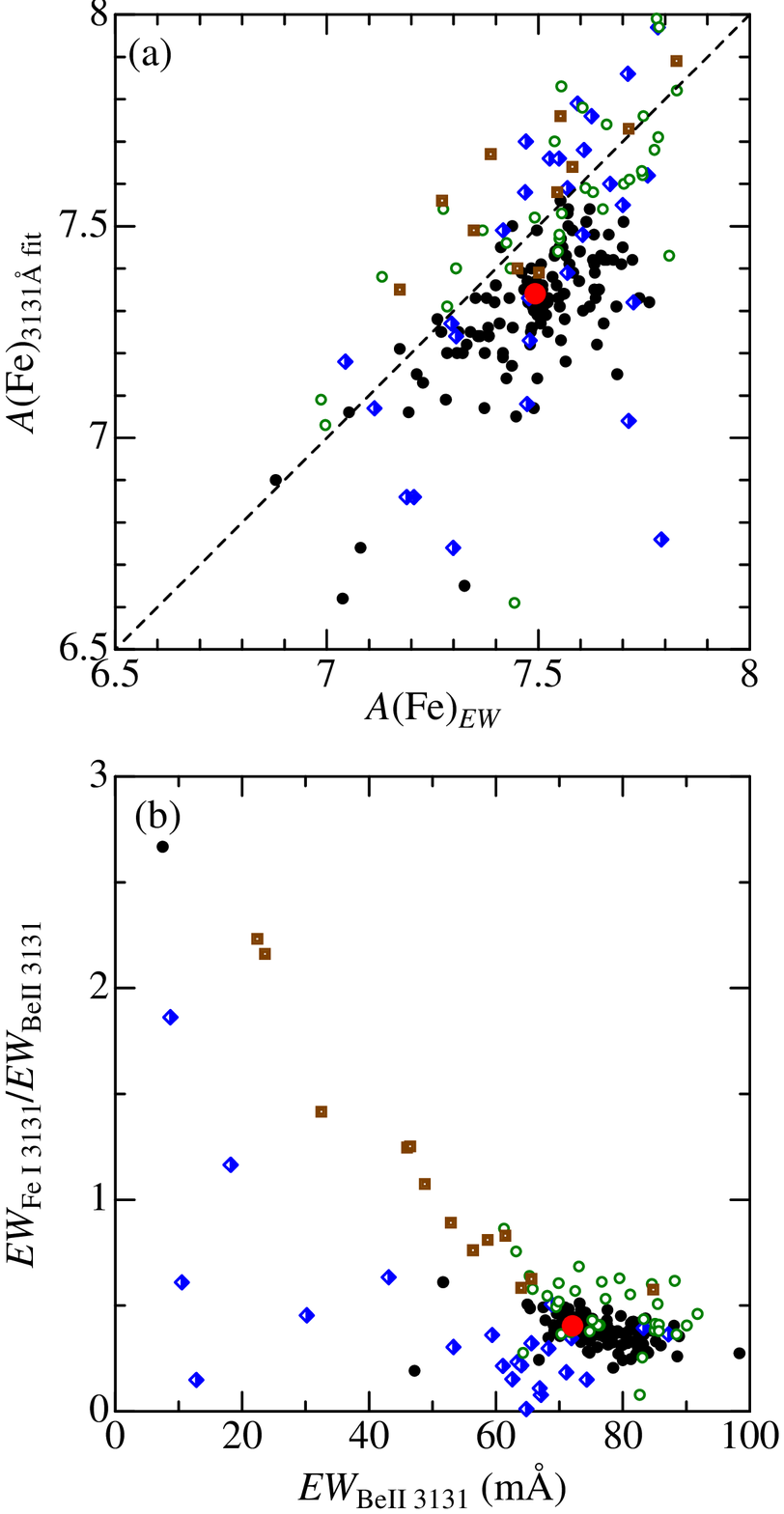}
  \end{center}
\caption{
(a) Correlation of the Fe abundance derived from the synthetic 
spectrum fitting analysis in the 3130.65--3131.35~$\rm\AA$ region 
with that determined based on $EW$s of many Fe lines (cf. Paper I,
Takeda et al. 2005). (b) Ratio of 
$EW_{\rm Fe I \;3131.043}/EW_{\rm Be II\; 3131.066}$
(equivalent widths inversely computed from fitting-based 
abundance solutions of Fe and Be) plotted against 
$EW_{\rm Be II \;3131.066}$.
The same meanings of the symbols as in figure 2.
}
\end{figure}

\subsection{Error Estimation}

By using the $EW$ values derived in the previous subsection, we tried 
to estimate errors involved in $A$(Be) from the viewpoints of two factors: 
ambiguities in the adopted atmospheric parameters and uncertainties 
in the treatment of line blending with the Be~{\sc ii} 3131.066 line.

The atmospheric parameters ($T_{\rm eff}$, $\log g$, $v_{\rm t}$) 
adopted in this study are the spectroscopically determined ones
taken from Paper I (solar analogs) and Takeda et al. (2005) (FGK
stars). These parameters were numerically derived (i.e., 
independently from any subjective judgement by human eye) based on the 
$EW$s measured for a number of Fe~{\sc i} and Fe~{\sc ii} lines,
while invoking the program TGVIT (Takeda et al. 2002, 2005) 
which establishes the best solutions satisfying (a) independence of 
$A$(Fe) upon $\chi_{\rm low}$, (b) independence of $A$(Fe) upon $EW$, 
and (c) matching of $\langle A$(Fe~{\sc i})$\rangle$ and 
$\langle A$(Fe~{\sc ii})$\rangle$. The intrinsic ambiguities
involved in these solutions may be defined as the extents of the
range, within which any change in the relevant parameter does not
cause any substantial influence on the judgement that these three
requirements (a)--(c) are fulfilled (see subsection 5.2 of Takeda et al. 
2002 for more details of the practical numerical procedures).
Such derived uncertainties in $T_{\rm eff}$, $\log g$, and $v_{\rm t}$
(along with the parameter solutions themselves) for all of the 
program stars are presented in electronic table E3.
As can be seen from these results, typical ambiguities\footnote{
The uncertainties quoted here are the ``internal'' statistical errors
being valid only in the relative sense within stars whose parameters 
were established in the same consistent manner.
We can not say here anything about systematic errors involved
in the absolute parameter values (e.g., those due to imperfectness
of the adopted atmospheric model, etc.).}
are on the order of 
$\sim \pm 20$~K in $T_{\rm eff}$, $\sim \pm 0.05$~dex in $\log g$, and 
$\sim \pm 0.1$~km~s$^{-1}$ in $v_{\rm t}$ (within a factor of $\ltsim 2$;
see also subsection 3.1 in Paper I and table E3 of Takeda et al. 2005). 
The Be abundance error corresponding to such estimated parameter 
uncertainties [$e(T_{\rm eff})$, $e(\log g)$, $e(v_{\rm t})$] for 
each star was evaluated as follows.
From the equivalent width of the Be~{\sc ii} line 
($EW_{\rm Be II \; 3131}$) described in the previous subsection, 
six kinds of abundances were derived while varying each of the standard 
values of atmospheric parameters interchangeably by $+e(T_{\rm eff})$, 
$-e(T_{\rm eff})$, $+e(\log g)$, $-e(\log g)$,
$+e(v_{\rm t})$, and $-e(v_{\rm t})$; and we call the 
differences of these perturbed abundances from the
standard one as $\delta_{T+}$, $\delta_{T-}$, $\delta_{g+}$, 
$\delta_{g-}$, $\delta_{v+}$, and $\delta_{v-}$, respectively.
We then computed the root-sum-square of three quantities
$\delta_{Tgv} \equiv (\delta_{T}^{2} + \delta_{g}^{2} + \delta_{v}^{2})^{1/2}$
as the abundance uncertainty (due to combined errors in 
$T_{\rm eff}$, $\log g$, and $v_{\rm t}$), 
where $\delta_{T}$, $\delta_{g}$, and $\delta_{v}$ are defined as
$\delta_{T} \equiv (|\delta_{T+}| + |\delta_{T-}|)/2$, 
$\delta_{g} \equiv (|\delta_{g+}| + |\delta_{g-}|)/2$, 
and $\delta_{v} \equiv (|\delta_{v+}| + |\delta_{v-}|)/2$,
respectively. The resulting values of 
$\delta_{T+}$, $\delta_{T-}$, $\delta_{g+}$, 
$\delta_{g-}$, $\delta_{v+}$, $\delta_{v-}$, and $\delta_{Tgv}$
for all the program stars are also given in electronic table 3.
An inspection of these results revealed, however, that $\delta_{Tgv}$ 
is only a few hundredths dex ($\sim$~0.02--0.06~dex) in most cases, 
which is mainly determined by $\delta_{g}$ (being larger than 
$\delta_{T}$ and $\delta_{v}$ by a factor of $\sim$~3--4). 
This indicates that the statistical errors in the atmospheric
parameters are practically insignificant for the results of
Be abundances.

Next, we tried to roughly estimate how the much errors in $A$(Be) may result 
from ambiguities in the contribution of the Fe~{\sc i} 3131.043 line. 
As mentioned in subsection 3.3 (figure 6a), the Fe abundances derived 
from the fitting analysis turned out more or less in accord with the 
[Fe/H] values resulting from $EW$s of a number of Fe lines as 
a by-product of parameter determinations (Paper I, Takeda et al. 2005).
Yet, an appreciable dispersion ($\sim \pm 0.2$~dex) seen in figure 6a
may be regarded as a measure of uncertainties in the treatment of
the blending effect. Accordingly, we proceeded as follows. 
We first derived $EW'_{\rm Fe I\; 3131}$ by using the fixed Fe abundance
corresponding to the [Fe/H] value (instead of the Fe abundance
obtained as the fitting solution) for each star, and then computed
the difference  
$\delta EW [\equiv EW'_{\rm Fe I\; 3131} - EW_{\rm Fe I\; 3131}]$.
Under the assumption that the sum of $EW_{\rm Be II\; 3131} + EW_{\rm Fe I\; 3131}$
remains invariable, we corrected $EW$ for the Be~{\sc ii} line as 
$EW'_{\rm Be II\; 3131} \equiv EW_{\rm Be II\; 3131} - \delta EW$,
from which the Be abundance ($A'$(Be)) corresponding to the case of using 
[Fe/H] of the star as given was calculated. Finally, the differences
between $A'$(Be) and $A$(Be) may be regarded as errors of $A$(Be) 
caused by uncertainties in the treatment of blending lines. 
(Here, the dispersion among these diferences is essentially 
important, since the absolute value of the difference itself 
depends on the error in the $gf$ value of the Fe~{\sc i} 3131.043 line.)
The resulting discrepancies of $\delta A$(Be) 
[$\equiv A'({\rm Be}) - A({\rm Be})$] computed for the program stars
are also given in electronic table 3, which revealed that 
$\delta A$(Be) mostly ranges from $\sim -0.3$ to $+0.1$ 
(dispersion of $\sim \pm 0.2$~dex) for FGK stars and 
$\sim -0.2$ to $\sim 0.0$ (dispersion of $\sim \pm 0.1$~dex)
for solar-analog stars. Accordingly, the corresponding errors
may be estimated as $\sim \pm 0.1$~dex and $\sim \pm 0.2$~dex 
for solar analogs and FGK stars, respectively.\footnote{
In a typical example case of $EW_{\rm Be II\; 3131} \sim$~70--90~m$\rm\AA$ 
and $EW_{\rm Fe I\; 3131} \sim$~30~m$\rm\AA$ for a 
solar analog, a perturbation of $\delta A$(Fe)~$\sim 0.2$~dex 
leads to $\delta EW_{\rm Fe I\; 3131} \sim$~7~m$\rm\AA$. This
change of $\delta EW \sim$~7~m$\rm\AA$ upon $EW_{\rm Be II\; 3131}$ 
($\sim$~70--90~m$\rm\AA$), in turn, varies $A$(Be) by $\sim 0.1$~dex.}
It should be remarked, however, that this discussion is valid
only for non-Be-depleted cases. As the strength of the Be line 
decreases, the error in $A$(Be) should be enhanced accordingly,
since the contribution of the Fe~{\sc i} line would become 
comparatively more important (actually a considerably large
$|\delta A$(Be)$|$ amounting to $\sim 1$~dex is even seen 
for a few Be-deficient cases). 

\subsection{Comparison with Previous Work}

The $A$(Be)'s resulting from this study and those derived by 
Santos et al. (2002, 2004a,b) and Boesgaard and Hollek (2009) for stars
in common are compared in figure 7. We can see from this figure
that Santos et al.'s results (mostly planet-host stars at 
6300~K~$\gtsim T_{\rm eff} \gtsim$~5400~K, tending to be metal-rich) 
are systematically lower than ours by $\sim$~0.1--0.3~dex, 
whereas Boesgaard and Hollek's results (for stars at 
6200~K~$\gtsim T_{\rm eff} \gtsim$~5800~K)
are slightly higher than ours by $\sim$~0.1--0.2~dex.
Regarding the comparison with the latter Boesgaard and Hollek's 
case, the appreciable discrepancies of $\sim 0.2$~dex shown by two 
Be-deficient stars ($A$(Be)~$\ltsim 1$; HD~89125 and HD~142860) may 
(at least partly) be attributed to the difference in the adopted 
atmospheric parameters, since the fact that their 
$T_{\rm eff}$/$\log g$ is lower/higher by $\sim 50$~K/$\sim 0.1$~dex 
than ours makes their $A$(Be) comparatively increase by $\sim 0.1$~dex. 
On the other hand, in the former case, we do not see any notable 
systematic differences in the adopted parameters between 
Santos et al. and ours, which suggests that the discrepancies 
may presumably stem from some difference in the $A$(Be) 
determination procedure.

\setcounter{figure}{6}
\begin{figure}
  \begin{center}
    \FigureFile(80mm,80mm){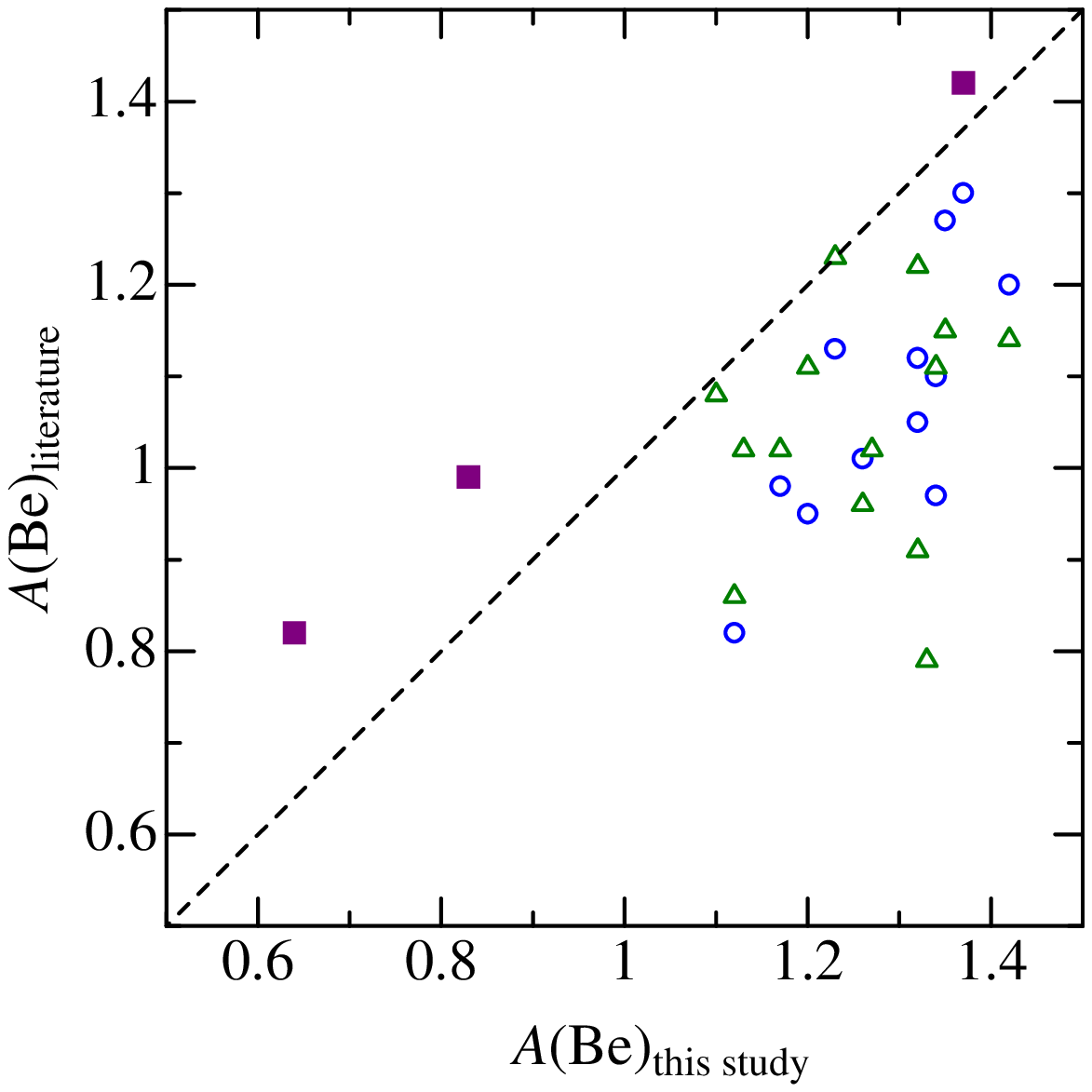}
  \end{center}
\caption{
Comparison of the beryllium abundances derived in this study
with the literature values: filled squares $\cdots$ Boesgaard 
and Hollek (2009; 3 stars in common); open circles $\cdots$
Santos et al. (2002; 13 stars in common); open triangles
$\cdots$ Santos et al. (2004a,b; 14 stars in common).
}
\end{figure}

\section{Discussion}

\subsection{General Behavior of Beryllium in FGK Stars}

Now that the Be abundances have been determined, we first discuss 
their general characteristics for the whole combined sample of 
205 stars (118 solar analogs and 87 FGK standards).

\subsubsection{Metallicity dependence of Be}

The $A$(Be) values obtained for all stars are plotted against 
[Fe/H] in figure 8b, where the results for $A$(Li) (Paper I, 
Takeda \& Kawanomoto 2005) are also shown (figure 8a) for comparison.
It is apparent from these figures that the trends are markedly
different between these two cases: while $A$(Li) are rather 
uniformly distributed over a considerably large span of $\gtsim 2$~dex, 
$A$(Be) tends to be confined in a narrow region between $\sim 1$
and $\sim 1.5$ (except for $T_{\rm eff} > 6000$~K stars showing
deficiencies).

The important problem to be clarified before discussing the
possibility of Be abundance variation is the primordial Be abundance
a star would have had when it was born. While it is known that
$A$(Be) and [Fe/H] are well correlated with each other in the 
metal-poor regime ([Fe/H]~$\ltsim -1$), the behavior for 
near-solar metallicity stars like the present sample 
($-0.5 \ltsim$~[Fe/H]~$\ltsim +0.4$) is still uncertain. 

Boesgaard et al.'s (2004a) analysis of solar-type dwarfs suggested
$\delta A$(Be)/$\delta$[Fe/H]~$\simeq 0.38$, whereas Boesgaard and 
Hollek (2009) recently argued (in their Be abundance study
of one solar-mass stars around $T_{\rm eff} \sim 6000$~K) that a 
steeper slope of $\delta A$(Be)/$\delta$[Fe/H]~$\simeq 0.86$ holds
not only for metal-poor stars but also for those up to 
[Fe/H]~$\sim 0$ continuously.
Meanwhile, there is a chemical evolution model that predicts a nearly 
constant (though rather complex) $A$(Be) at  $-1 \ltsim$~[Fe/H] 
(cf. figure 4 of Casuso \& Beckman 1997).

The upper envelope of our $A$(Be) vs. [Fe/H] relation in figure 8b
indicates that $A$(Be) certainly increases with [Fe/H].
According to the linear-regression analysis applied to solar analogs
(mostly with $1.0 \le$~$A$(Be)~$\le 1.5$, while excluding several outliers) 
at $-0.3 \ltsim$~[Fe/H]~$\ltsim +0.3$, we obtained
$A$(Be) = 1.28 ($\pm 0.01$) +0.49 ($\pm 0.05$) [Fe/H], which suggests
$\delta A$(Be)/$\delta$[Fe/H]~$\sim 0.5$ (i.e., just the midpoint 
between 0 and 1).
Accordingly, when showing the trend of Be for discussing
the possibility of depletion-induced abundance changes, we present 
two kinds of figures in terms of $A$(Be) as well as [Be/Fe],
considering that the relevant information may be revealed by these two.

\subsubsection{Abundance trend with $T_{eff}$}

Figures 9a and b display the distribution of $A$(Be) and [Be/Fe]
against $T_{\rm eff}$, from which we can notice the following 
characteristics:\\ 
--- For F-type stars ($T_{\rm eff} > 6000$~K), Be abundances show
a considerably large spread amounting to $\sim 2$~dex, as has been 
already reported in previous studies (e.g., Boesgaard et al. 2004b).\\ 
--- Regarding early-G stars at 6000~K~$> T_{\rm eff} >$~5500~K,
the main targets of this study, we can recognize that most stars have
similar Be abundances near to the solar value, but there {\it do} exist 
several stars showing conspicuous Be depletion. Interestingly,
all these stars have almost the same $T_{\rm eff}$ as the Sun.
Accordingly, we may state that the existence of ``Be-gap'' 
(i.e., a marked depression of Be abundances in a narrow $T_{\rm eff}$ range) 
suggested by Santos et al. (2004a) has been confirmed (in a limited sense) 
in our solar-analog sample.
The origin of Be depletion for these stars will be discussed 
in subsection 4.2.\\
--- As to late-G or early-K stars (5500~K~$> T_{\rm eff} >$~5000~K), 
Be is definitely depleted for three lowest $T_{\rm eff}$ stars 
($T_{\rm eff} < 5250$~K), 
which is understandable because they are rather evolved subgiants 
(cf. figure 2b) and envelope mixing is considered to be enhanced. 
If we restrict to dwarfs, there may be a weak signature 
that depletion gradually begins at $T_{\rm eff} \ltsim $~5500~K.\\
--- We could not confirm the decreasing tendency of $A$(Be) with
a lowering of $T_{\rm eff}$ reported by Santos et al. (2004a,b).
Rather, the upper envelope of $A$(Be) (and especially of [Be/Fe]) 
appears to slightly increase with a decrease in $T_{\rm eff}$
from $T_{\rm eff} \sim 6500$~K down to $T_{\rm eff} \sim 5500$~K.
Actually, this is a tendency already suggested by Boesgaard and 
Hollek (2009; cf. their figure 9).

\setcounter{figure}{8}
\begin{figure}
  \begin{center}
    \FigureFile(80mm,80mm){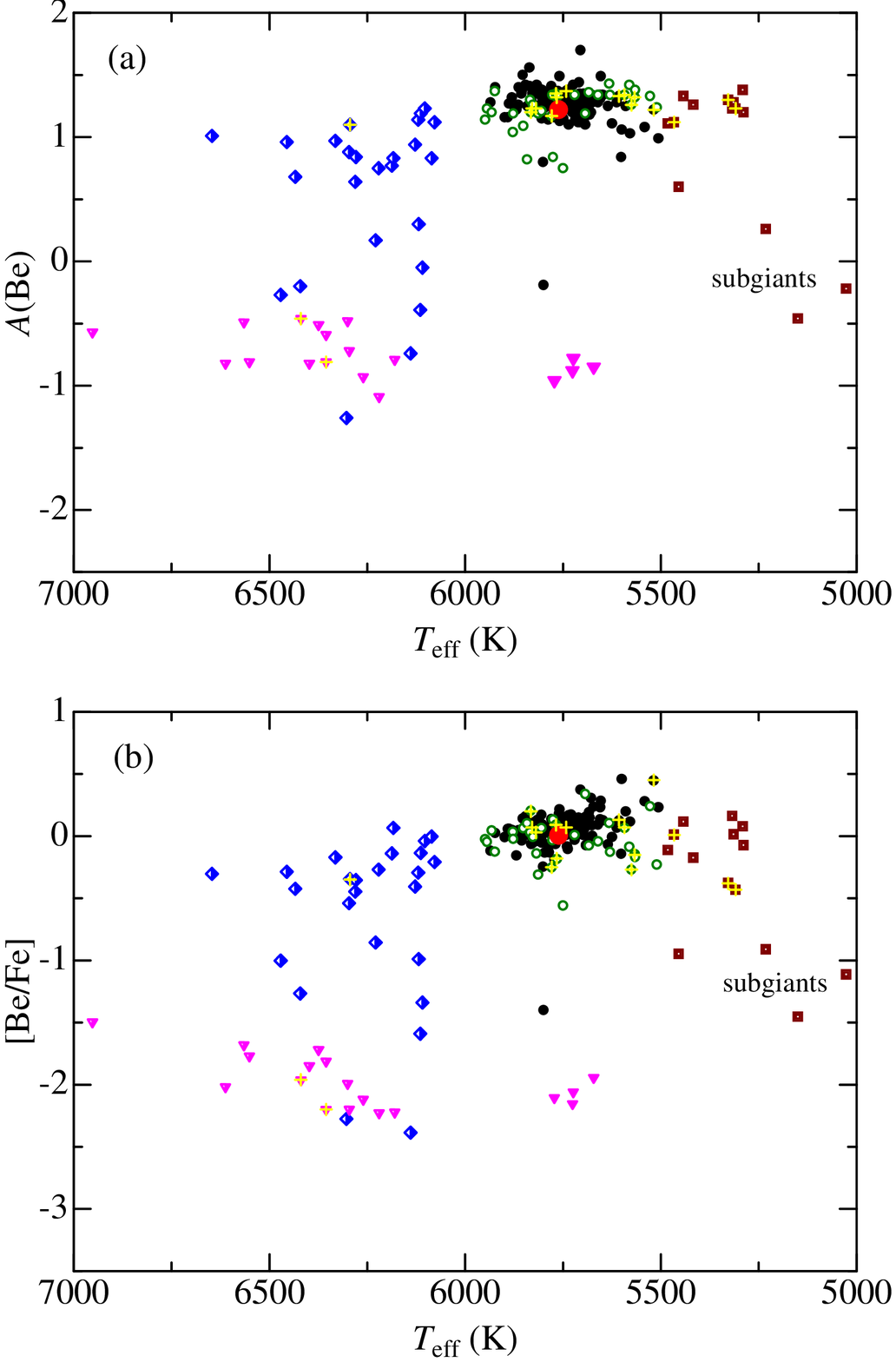}
  \end{center}
\caption{
Beryllium abundances plotted against the effective temperature.
(a) $A$(Be) vs. $T_{\rm eff}$; (b) [Be/Fe] vs. $T_{\rm eff}$.
The same meanings of the symbol types as described in the caption
of figure 2. The downward triangles (pink) denote the upper-limit
values for the indeterminable cases where the Be~{\sc ii} 3131.066 
line is too weak to be detectable.
The planet-host stars are indicated by overplotting yellow crosses.
}
\end{figure}

\subsubsection{Beryllium in planet-host stars}

It is worthwhile to examine whether there is any difference in the 
behavior of Be between planet-host stars (PHS) and non-planet-host 
stars (non-PHS), since our sample includes 18 such stars harboring 
giant planets (5 in the solar analogs and 13 in the FGK standards), 
the Be abundances of which are summarized in table 2.
As seen from this table, the $A$(Be) values for most (sixteen) PHS are 
around the solar value ($\sim 1.2 \pm 0.2$) to within $\ltsim 0.2$~dex 
similar to other solar-type non-PHS, while only two PHS at 
$T_{\rm eff} \sim 6400$~K show a marked depletion as other F-type 
non-PHS. Accordingly, we may state that any meaningful difference 
can not be noticed with regard to the surface Be abundances between 
two samples of stars with and without giant planets (which can be 
confirmed by plotting the data of these 18 stars on figure 9).
This result is essentially a reconfirmation of the conclusion of 
Santos et al. (2002, 2004b, 2010).
In this connection, it is worth pointing out that HD~186408 
(16~Cyg~A, non-PHS, $A$(Be) = 1.34, $A$(Li) = 1.37)
and HIP~96901 (16~Cyg~B, PHS, $A$(Be) = 1.37, $A$(Li)~$<1.06$ ), 
quite similar Sun-like non-PHS and PHS stars constituting a visual binary 
system, have practically the same $A$(Be) despite the distinct difference 
in $A$(Li).

\subsection{Be Abundances in Solar-Analog Stars}

\subsubsection{Evidence of rotation-dependence}

We now confine ourselves to the main topic of this study:
How is the behavior of the Be abundances of 118 solar-analogs?
Do they have any correlation with Li abundances or other stellar
properties such as rotation or age?

We can see several interesting characteristics in figure 10, where
the results of $A$(Be) and [Be/Fe] are plotted against 
$A$(Li), $v_{\rm e}\sin i$, $r_{0}$(8542), and $\log age$.\\
--- While most stars have similar Be abundances around the solar
value within $\ltsim 0.2$~dex, 4 stars are markedly discrepant
from the general trend of $A$(Be), for which the Be~{\sc ii} 3131 line 
is undetectable, suggesting a drastic Be depletion by more than 
$\sim 2$~dex: HIP~17336 ($A$(Be)$<-0.85$), HIP~32673 ($<-0.78$), 
HIP~64150 ($<-0.88$), and HIP~75676 ($<-0.96$).\footnote{ 
HIP~88945 with $A$(Be) = $-0.19$ should also be classified 
as a Be-depleted star. However, as already remarked in Paper II,
this star has rather unusually high $v_{\rm e}\sin i$ and 
$r_{0}$(8542) in spite of its low $A$(Li), which means
that it does not follow the general trend of 
$A$(Li)--$v_{\rm e}\sin i$--$r_{0}$(8542) relation established 
in Paper II. Therefore, we should be cautious in interpreting
the results for this outlier star.}
It is important to note that, in all these Be-depleted stars, 
(i) Li is also depleted, (ii) $v_{\rm e}\sin i$ is the lowest class, 
and (iii) $r_{0}$(8542) is also the lowest class (that is, these 
4 stars are clustered in the lower-left corner in figures 10a--c
as well as a$'$--c$'$). This indicates that slow rotation must be
the key factor (at least one of the causes, even if not the only one) 
triggering such an exceptionally drastic dearth of Be in solar-analogs.\\
--- Furthermore, regarding the other stars (and the Sun) having 
apparently similar Be abundances, we notice from figures 10a--c
as well as a$'$--c$'$ a weak but significant trend that $A$(Be) 
(and also [Be/Fe], though less clearly) shows a gradual rise with 
increasing $A$(Li), $v_{\rm e}\sin i$, and $r_{0}$(8542).
This may be interpreted that some kind of very gradual Be depletion 
mechanism takes place even in these ``superficially normal Be stars,'' 
acting more efficiently as stellar rotation becomes slower.

\subsubsection{What are Be-depletion mechanisms like?}

Considering these two facts, we may conclude that stellar rotation
plays a significant role in controlling the surface Be abundances
of solar-analog stars. However, it is rather hard to imagine that 
both the drastic Be depletion for a tiny fraction of stars and 
the more gradual alteration for the majority of stars are 
attributed to the same physical process.
Accordingly, we would propose that two kinds of Be-depletion 
mechanism may operate in the envelope of Sun-like stars; 
i.e., the ``strong'' and ``weak'' ones: \\
--- The ``strong'' process should work only in limited cases 
where special conditions are satisfied; but once it has been triggered, 
surface Be is depleted very efficiently. At least, very slow rotation 
(and presumably also the near-solar $T_{\rm eff}$) 
may be counted as the necessary (if not sufficient) condition 
for this process to work, though we have no specific idea which kind of 
physical process it is.\footnote{
An anonymous referee kindly suggested a possibility
that these four conspicuously Be-deficient stars might belong to a group 
of stars having rotated very rapidly in the past (and suffered substantial
Be depletion due to rotation-induced mixing) but considerably spun down 
at present, as suggested from the modeling of ultra-Li-deficient halo stars
(Pinsonneault et al. 1999, 2002), which is an intriguing hypothesis.
However, we should bear in mind that these 4 stars have parameters 
quite similar to the Sun (actually, 3 are the solar twin candidates 
sorted out in Table A.1 of Paper I), also with respect to $v_{\rm e}\sin i$. 
So, if this is really the case, we should regard that the rotational 
braking of these stars must have been so strong that any trace of 
having being a rapid rotator in the past has completely disappeared.
}
\\
--- On the other hand, the ``weak'' process is relevant for most of 
the stars, steadily and slowly acting to reduce Be in the outer 
envelope, which should be more efficient as the rotation becomes slower.
Such a mechanism is likely to be the bit-by-bit destruction of
Be nuclei at the bottom of the convection zone (where the Be 
burning temperature of $\sim 3.5 \times 10^{6}$~K is not yet 
reached; cf. figure 1) by an appropriate mixing of the boundary layer
caused by differential rotation-induced shear such as that 
considered by Bouvier (2008) (cf. subsection 1.2),
since this theory most reasonably explains the observational fact
among several possible $A$(Be)-changing mechanisms investigated 
so far.\footnote{
For example, Deliyannis and Pinsonneault (1997) 
theoretically studied the effect of mass loss, diffusion, and 
wave-driven as well as rotational mixing in the envelope of late F stars.
Similarly,  Charbonnel and Lagarde (2010) recently examined the 
role of rotation-induced mixing along with the thermohaline instability 
for the Li depletion in the envelope of 1--4~$M_{\odot}$ stars.
However, the conclusions of these studies suggest that faster rotation 
enhances the mixing, which is just the opposite to what we found here.
}
\\
--- In cases where such a long-lasting ``weak'' depletion mechanism
takes place, it is expected that older stars show larger Be-deficiency.
Unfortunately, we can not make a decisive answer about this test, 
since $A$(Fe) and [Be/Fe] exhibit different behaviors due to
the tight [Fe/H] vs. $age$ relation (figure 2c); i.e., while 
$A$(Be) certainly shows this expected trend (figure 10d), the 
tendency of [Be/Fe] is just the opposite (figure 10d$'$).
In this connection, we should also pay attention to the possibility
that mixing may be rather suppressed in old stars, because they are
generally of low metallicity which makes convection less active.

\subsubsection{Is solar Be abundance primordial?}

The consequence from figure 10, that any solar-analog stars must 
have (more or less) suffered  Be-depletion in the envelope 
with its extent depending on the rotational velocity, naturally
implies that most stars (especially slowly rotating ones) 
currently have surface Be abundances appreciably lower than 
the primordial values they had at their birth. 
Then, since our Sun definitely belongs to the slow-rotation group
($v_{\rm e}\sin i \sim 2$~km~s$^{-1}$), a significant discrepancy 
between the current and the initial Be abundances is expected.
We may estimate this amount as $\sim$~0.2--0.3~dex from the abundance 
difference between $v_{\rm e}\sin i \sim 2$~km~s$^{-1}$
and $v_{\rm e}\sin i \sim 10$~km~s$^{-1}$ in figures 10b or b$'$,
assuming that the depletion is sufficiently small for the latter 
high-rotation case. Interestingly, this value quite reasonably 
explains the observed difference between the current photospheric 
Be abundance (1.22; from this study) and the meteoritic Be abundance 
(1.42;  Anders \& Grevesse 1989).
As a matter of fact, this meteoritic--photospheric discordance of Be
was previously regarded as due to a mixing-induced depletion 
in the solar envelope.

However, Balachandran and Bell (1998) argued against this scenario,
insisting that this discrepancy simply stemmed from erroneous
underestimation of the solar Be abundance due to imperfect
treatment of UV continuum opacity (so-called ``missing opacity'') 
at the Be~{\sc ii} line region.
They showed that a photospheric abundance consistent with the
meteoritic value is obtained by increasing the continuum opacity 
by a factor of 1.6. Based on this argument, they concluded
that the mixing in the solar envelope is too shallow to affect
the Be abundance at the surface, in which the primordial (meteoritic)
abundance of $A$(Be)$\sim 1.4$ is retained unchanged.
Furthermore, Asplund (2004) corroborated their conclusion based on
the discussion of oxygen abundance matching between different lines,
though it may not be regarded as a direct evidence for
the existence of missing opacity.

Yet, Balachandran and Bell's (1998) 
argument conflicts with our interpretation mentioned above.
Therefore, we decided to examine by ourselves whether such a UV 
missing opacity problem really exists around the Be~{\sc ii} 3131 
line region. The comparison of the observed solar flux energy distribution
at the earth (i.e., just outside of earth's atmosphere) taken from
Woods et al. (1996) with the theoretical flux computed from Kurucz's (1993) 
ATLAS9 solar model atmosphere (reduced to the value at the earth\footnote{
$f_{\lambda}^{\rm th} \equiv 4 \pi H_{\lambda}^{\rm th}(0) (R/d)^{2}$,
where $H_{\lambda}^{\rm th}(0) [\equiv 
\frac{1}{2} \int_{0}^{1} \mu I_{\lambda}^{\rm th}(0, \mu) d\mu]$ 
is the Eddington flux at the surface of the Sun, $R$ is
the solar radius, and $d$ is the Sun--earth distance.})
is shown in figure 11. Rather unexpectedly, as seen from this figure, 
we could not confirm any such significant flux discrepancy between 
theory and observation at $\lambda \sim 3131\; \rm\AA$ that may suggest 
an existence of missing opacity in the model.

\setcounter{figure}{10}
\begin{figure}
  \begin{center}
    \FigureFile(80mm,140mm){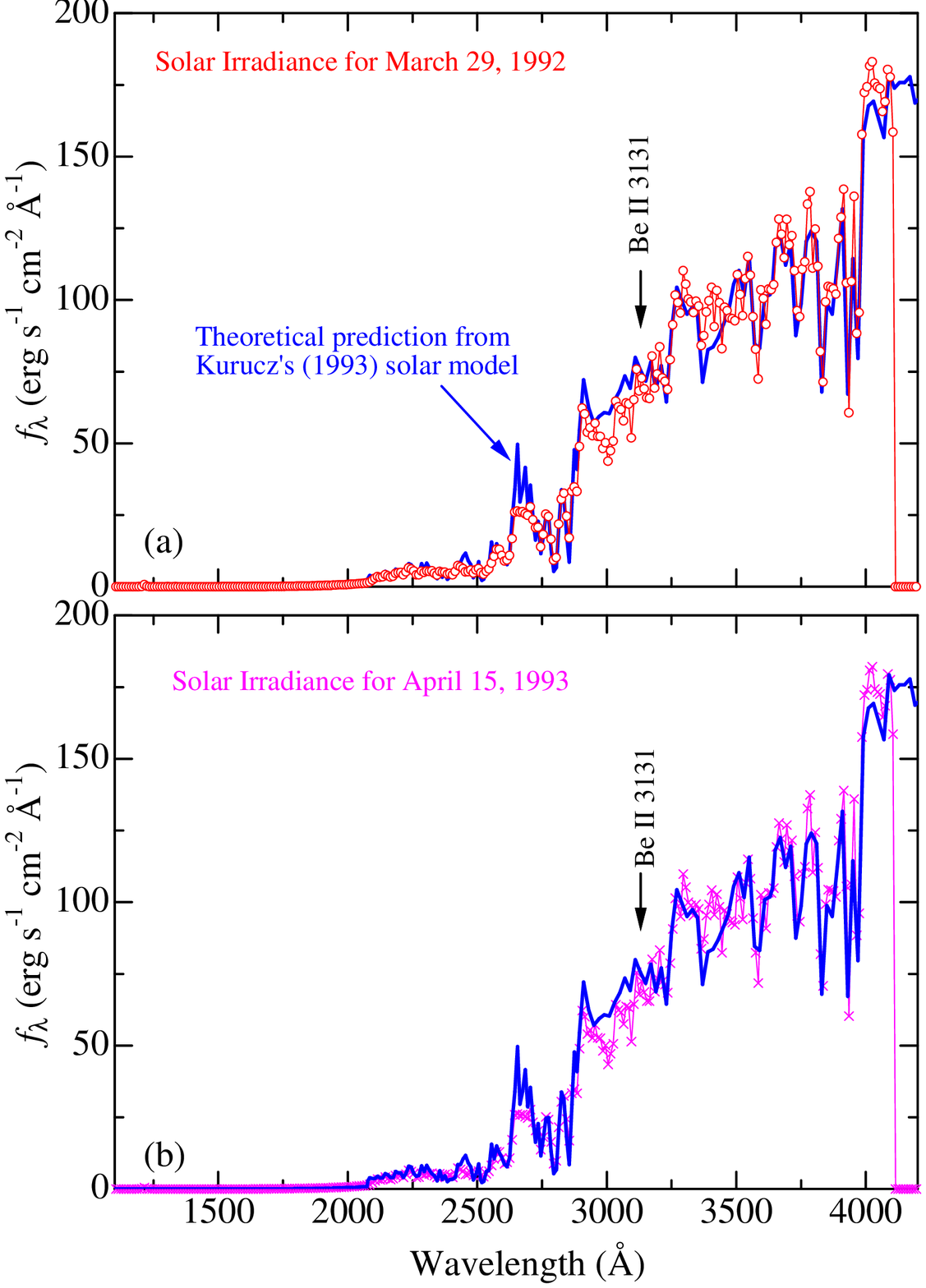}
  \end{center}
\caption{
Comparison of the absolute solar flux at the position of the earth 
in the 1100--4100~$\rm\AA$ region. Theoretical fluxes computed with
Kurucz's (1993) ATLAS9 solar model atmosphere (reduced to the values
at the earth) are depicted by thick lines, while the observed data 
of Woods et al. (1996) on two different dates are shown by 
line-connected symbols. 
(a) Experiment on March 29, 1992 (open circles);
(b) Experiment on April 15, 1993 (crosses).
}
\end{figure}

This result casts doubt about the reliability of the conclusion made by 
Balachandran and Bell (1998), in which we actually notice some 
questionable points:\\
--- They appear to put too much weight on the delicate difference
seen in the profiles of spectral line at the $\lambda \sim 3128 \; \rm\AA$ 
and $\lambda \sim 3131 \; \rm\AA$ region in order to justify
their claim, stating that adding an extra amount of UV opacity 
significantly improves the fit between theoretical and observed line 
shapes. In our opinion, however, solution to such a line-profile problem 
may be found in some other ways, such as adjusting the line-broadening 
function due to macroturbulence (known to be depth-dependent in the
solar atmosphere).\\
--- If confirming the existence of missing opacity is the central issue,
modeled absolute flux distribution should be compared with the observed
solar ultraviolet irradiance as we did in figure 11.
We would point out, however, that any direct evidence of 
discrepancy between theoretical and observational solar fluxes just at 
$\lambda \sim 3131\; \rm\AA$ has never been presented, as far as we see
the papers by Gustafsson and Bell (1979) or Bell, Paltoglou, and 
Trippico (1994), which Balachandran and Bell (1998) quoted for 
the indication of too large calculated solar flux.

Of course, it is premature to conclude that their consequence is 
incorrect based on this argument alone; further check or reconfirmation 
may be required.\footnote{
In this connection, Smiljanic et al. (2009) reported 
that an increase in the Fe~{\sc i} bound-free opacity by a factor 
of 1.6 (the enhancement factor proposed by Balachandran and Bell)
in a test model with [Fe/H] = $-0.5$ hardly affects the resulting 
Be abundance in the practical sense (i.e, only 0.022~dex),
suggesting that UV missing opacity is insignificant.} 
Yet, at any event, given that the observed solar UV 
flux is well reproduced by the ATLAS9 model, which we also invoked for
Be abundance determinations, there is no reason for us to concern 
about any missing opacity. We may thus regard that $A_{\odot}$(Be)~=~1.22
derived in this study represent the ``real'' beryllium abundance
in the solar photosphere, and that its being lower by 0.2~dex compared to
the meteoritic abundance is attributed to the result of a mild depletion 
in the solar envelope. In short, the currently observed Be abundance 
in the photosphere of the Sun (and in solar-analog stars) should not 
be regarded as representing the primordial abundance, which must have 
gradually varied during the lifetime so far.

\section{Conclusion}

We conducted an extensive beryllium abundance analysis for 118 
solar analogs (along with 87 FGK standard stars) in order to
clarify the controversial issue of how the Be abundances for Sun-like 
stars behave themselves at 6000~K~$\gtsim T_{\rm eff} \gtsim$~5500~K
in connection with the lithium abundance and stellar parameters 
such as rotation, age, and activity index. In particular, we wanted
to answer the question ``Do $A$(Be)'s show any sign of depletion 
like $A$(Li)? Or, alternatively, are they almost 
constant at the primordial Be abundances unchanged?''

The observed spectra of 205 targets in the near-UV region including 
Be~{\sc ii} resonance lines were obtained in the 2009--2010 season 
by using the High-Dispersion Spectrograph of the Subaru Telescope.
The Be abundance of each star was determined by applying the spectrum 
synthesis technique (coupled with the automatic best-fit solution
finding algorithm) to the 3130.65--3131.35~$\rm\AA$ region comprising
the Be~{\sc II} line at 3131.066~$\rm\AA$.

Inspecting the trend of the Be abundances for the whole sample, 
we found that $A$(Be)'s for a majority of stars (especially for 
the solar-type) are confined around the solar value, though 
stars showing conspicuous Be depletion do exist (especially for 
F-type stars at $T_{\rm eff} \gtsim$~6000~K). This makes 
a marked contrast to the case of $A$(Li) which tends to show a rather 
uniform spread with a considerably large dispersion of $\gtsim 2$~dex.
The upper envelope of $A$(Be) vs. [Fe/H] relation suggests
that $A$(Be) increases with [Fe/H] with a slope of 
$\delta A$(Be)/$\delta$[Fe/H]~$\sim 0.5$ in the metallicity range of
$-0.5 \ltsim$~[Fe/H]~$\ltsim +0.4$. We could not find any 
difference in Be between stars with and without planets.

Regarding the Sun-like stars, while most of them are superficially 
similar in terms of their $A$(Be) at $\sim 1.2 \pm 0.2$, 4 out of 
118 stars turned out strikingly Be-deficient (by $\gtsim 2$~dex) 
and these 4 stars have the lowest $v_{\rm e}\sin i$, 
the lowest stellar activity, and considerably depleted $A$(Li). 
Moreover, even for the other majority showing an apparent 
similarity in Be, we can recognize a tendency that $A$(Be) 
gradually increases with an increase in $v_{\rm e}\sin i$, 
$A$(Li), and $r_{0}$(8542).

These results suggest that any solar analog star (including 
the Sun) generally suffers some kind of rotation-dependent 
Be depletion, for which we suspect two kinds of mechanisms 
may operate: The ``strong'' process should work only in limited 
cases under special conditions but depletes surface Be 
very efficiently once triggered, whereas the ``weak'' process 
acts on most stars and slowly reduce Be in the outer envelope.
Contributions of theoreticians are desirably awaited toward 
developing a reasonable model/theory accounting for the 
observational facts.

According to our findings, the occasionally stated view 
``Be-depletion is irrelevant to G-type dwarfs with 
$T_{\rm eff} \ltsim 6000$~K, which have similar Be abundances 
and retain their original composition'' is not correct; 
Be abundances of such stars in general are considered to have 
more or less suffered reduction compared to the primordial values. 
The difference of $\sim 0.2$~dex between the current solar 
photospheric Be abundance and the meteoritic Be abundance may be 
interpreted in this way. This means that we do not lend 
support for the widely mentioned scenario of ``erroneous 
underestimation of Be abundances due to the neglect of 
UV missing opacity.''

\bigskip

This research has made use of the SIMBAD database, operated by
CDS, Strasbourg, France. 

\appendix

\section*{Non-LTE Effect on the Be II 3131 Line}

The non-LTE effect on the formation of Be~{\sc ii} 3130--3131
doublet has already been investigated by several investigators
(see, Asplund 2005 for a review of the previous work), such as 
Chmielewski, M\"{u}ller and Brault (1975) [for the Sun], 
Shipman and Auer (1979) [for the Sun], 
Kiselman and Carlsson (1995) [for the Sun, Procyon, and HD~140283],
and Garc\'{\i}a L\'{o}pez, Severino, and Gomez (1995) [for the Sun and 
metal-poor stars]. All these performed non-LTE calculations on the 
Be~{\sc i} + Be~{\sc ii} atomic model and arrived at almost the same 
conclusion that the non-LTE correction for deriving the Be abundance 
from the Be~{\sc ii} 3130--3131 resonance lines is insignificant.

Since FGK stars covering a rather wide $T_{\rm eff}$ range
(7000~K~$\gtsim T_{\rm eff} \gtsim$~5000~K) are concerned
in the present study, and we are interested in how the treatment 
of neutral-hydrogen collisions influences the results
(which has not been considered in previous studies
of Be line formation), we carried out some test calculations.

Our atomic model of beryllium comprises 20/22 terms and 28/89 
radiative transitions for Be~{\sc i}/Be~{\sc ii}, including up to 
Be~{\sc i} 2$s$\,6$p$~$^{1}$P (71746~cm$^{-1}$ from the ground level) 
and Be~{\sc ii} 7$g$~$^{2}$G (137925~cm$^{-1}$ from the ground level),
which was constructed from the atomic-line database compiled 
by Kurucz and Bell (1995).
The data from TOPbase (Cunto \& Mendoza 1992) were adopted for 
the photoionization cross sections for the lowest 9 and 8 terms of 
Be~{\sc i} and Be~{\sc ii}, respectively. As to other computational
details (e.g., electron-collision rates, 
photoionization rates for the remaining terms, 
collisional ionization rates,  treatment of collisions
with neutral-hydrogen atoms, etc.), we followed the recipe 
described in subsubsection 3.1.3 of Takeda (1991).

The statistical-equilibrium calculations were carried out for 
four models with $T_{\rm eff}$ = 5000, 5500, 6000, and 6500~K
(each with the same solar-metallicity and $\log g=4$).
Regarding the treatment of neutral-hydrogen collision rates,
we tested two cases: (i) classical H~{\sc i} rates are used 
uncorrected ($k=1$), (ii) classical H~{\sc i} rates are drastically 
reduced by multiplying them by a factor of $k=10^{-3}$.
 
In figure 12 are shown the $S_{\rm L}(\tau)/B(\tau)$ (the ratio of 
the line source function to the Planck function, and nearly equal to 
$\simeq b_{\rm u}/b_{\rm l}$, where $b_{\rm l}$ and $b_{\rm u}$ are 
the non-LTE departure coefficients for the lower and upper levels, 
respectively) and  $l_{0}^{\rm NLTE}(\tau)/l_{0}^{\rm LTE}(\tau)$ 
(the NLTE-to-LTE line-center opacity ratio, and nearly equal to 
$\simeq b_{\rm l}$) for the multiplet 1 transition
(2$s$\,$^{2}$S--2$p$\,$^{2}$P$^{\circ}$) relevant to the 
Be~{\sc ii} 3130--3131 doublet lines.
Also, the LTE and non-LTE equivalent width for the 
Be~{\sc ii} 3131.066 line and the corresponding non-LTE abundance
correction were computed for each case, which are summarized in table 3.

Inspecting figure 12 and table 3, we can make the following conclusions
regarding the non-LTE effect on the Be~{\sc ii} 3131.066 line:\\
--- While an appreciable departure from LTE is noticed in the line
opacity (increased opacity acting to strengthen absorption) 
as well as in the line source function (enhanced source function
acting to dilute/weaken absorption), the net effect on the equivalent
width is quite small, because both act in the opposite direction
(see, e.g.,  Asplund 2005 or Garc\'{\i}a L\'{o}pez et al. 1995a
for more details concerning the line-formation mechanism).\\ 
--- The results are practically independent of how the H~{\sc i}
collision is treated, which reflects that the formation
of this UV transition is controlled by (not collisional but)  
radiative processes.\\
--- We can neglect the non-LTE abundance correction (especially
for solar-type stars with $T_{\rm eff} \ltsim 6000$~K), since
its extent is on the order of a few hundreds dex at most, 
though it may become noticeable for F-type stars of
$T_{\rm eff} \gtsim 6500$~K at a level of $\sim 0.1$~dex
(positive correction).

\setcounter{figure}{11}
\begin{figure}
  \begin{center}
    \FigureFile(80mm,140mm){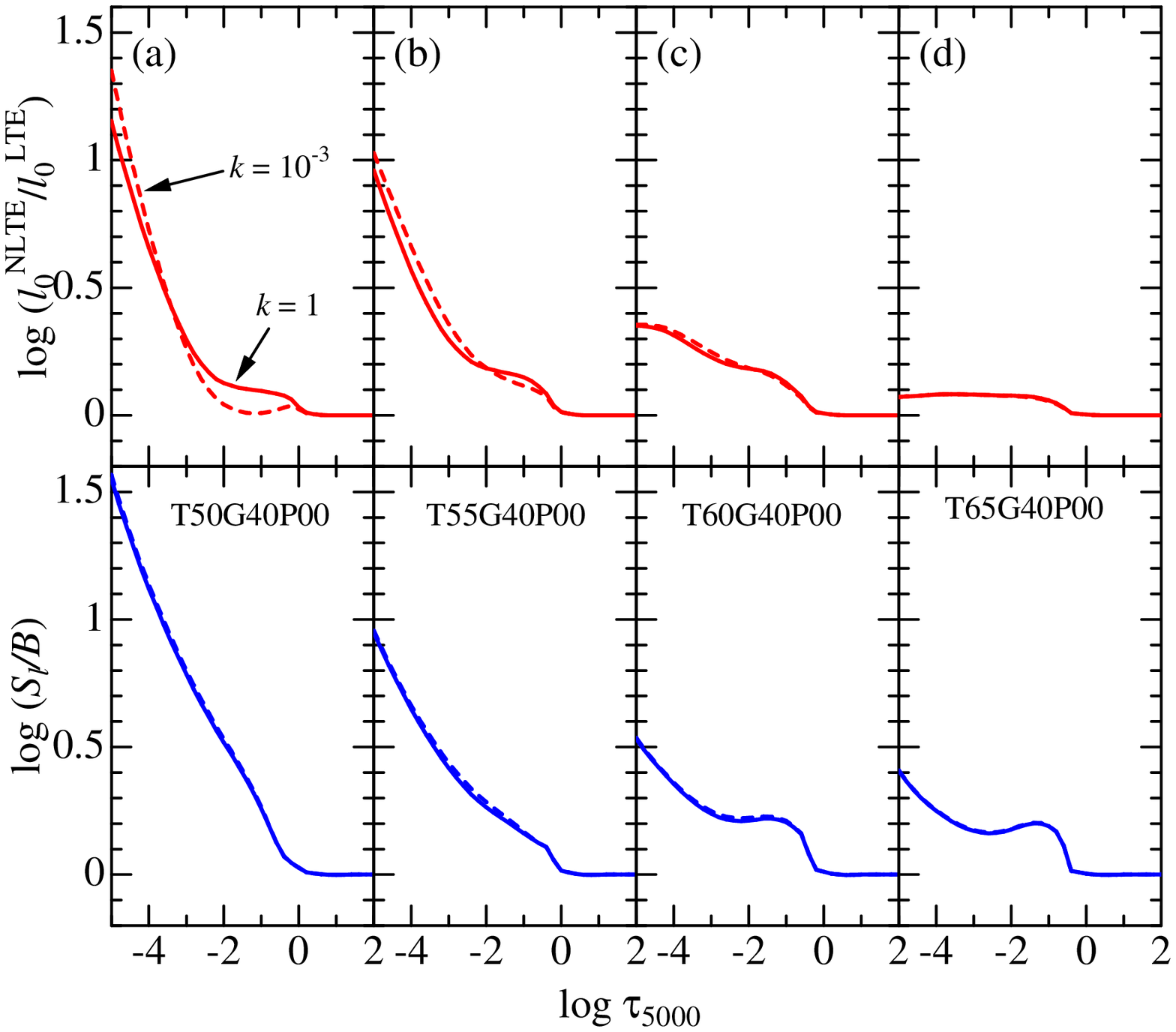}
  \end{center}
\caption{
Departure from LTE for the Be~{\sc ii} 2$s$\,$^{2}$S--2$p$\,$^{2}$P$^{\circ}$ 
transition (multiplet 1) corresponding to the Be~{\sc ii} lines at
3130.421~$\rm\AA$ and 3131.066~$\rm\AA$, computed for 
four representative atmospheric models 
with different $T_{\rm eff}$ but with the same $\log g$ (4.0) 
and the metallicity (solar). The upper and lower panels show the 
depth-dependence of the NLTE-to-LTE line-center opacity ratio 
and the ratio of the line source function to the Planck function,
respectively. Panels (a), (b), (c), and (d)
correspond to $T_{\rm eff}$ of 5000~K, 5500~K, 6000~K, and
6500~K, respectively. Solid lines $\cdots$ results for $k=1$
(classical treatment of the H~{\sc i} collision without any 
correction); dashed lines $\cdots$ results for $k = 10^{-3}$ 
(reduction of the classical H~{\sc i} collision rates by 
a factor of $10^{-3}$; i,e., practically to a negligible level).
}
\end{figure}


\newpage


\onecolumn


\setcounter{figure}{3}
\begin{figure}
  \begin{center}
    \FigureFile(170mm,220mm){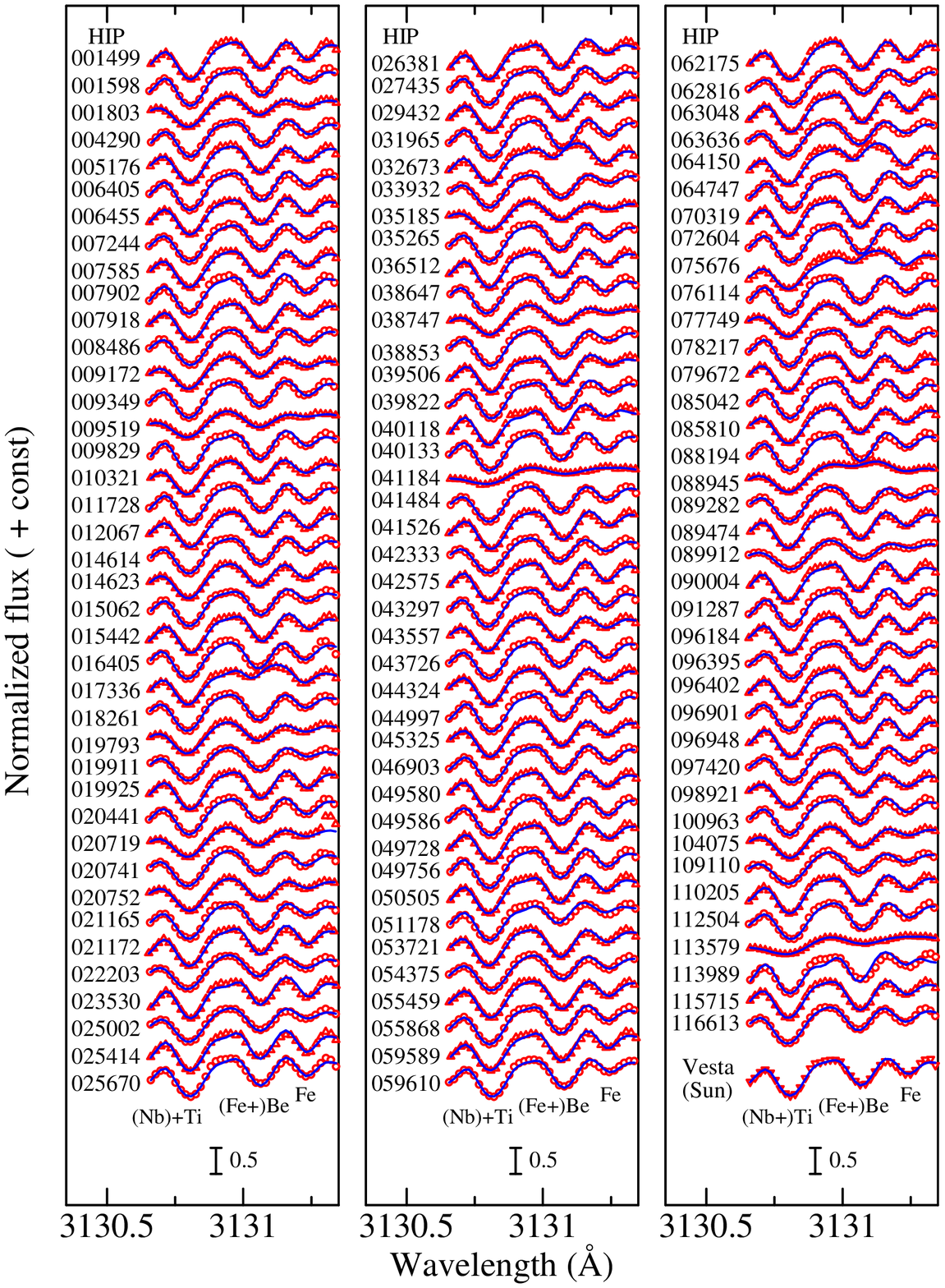}
  \end{center}
\caption{
Synthetic spectrum fitting in the 3130.65--3131.35~$\rm\AA$ region 
for 118 solar-analog stars. The best-fit theoretical spectra
are shown by solid lines, while the observed data are plotted
by symbols. A vertical offset of 0.5 is applied to each relative to
the adjacent ones. Each of the spectra are arranged in the
increasing order of HIP number (indicated on the left to each 
spectrum), as in figures 6/8 in Paper I or figure 2 in Paper II, 
while the case of Vesta (Sun) is displayed at the bottom-right.
The wavelength scale of each spectrum is adjusted to the 
laboratory system.
}
\end{figure}

\setcounter{figure}{4}
\begin{figure}
  \begin{center}
    \FigureFile(170mm,220mm){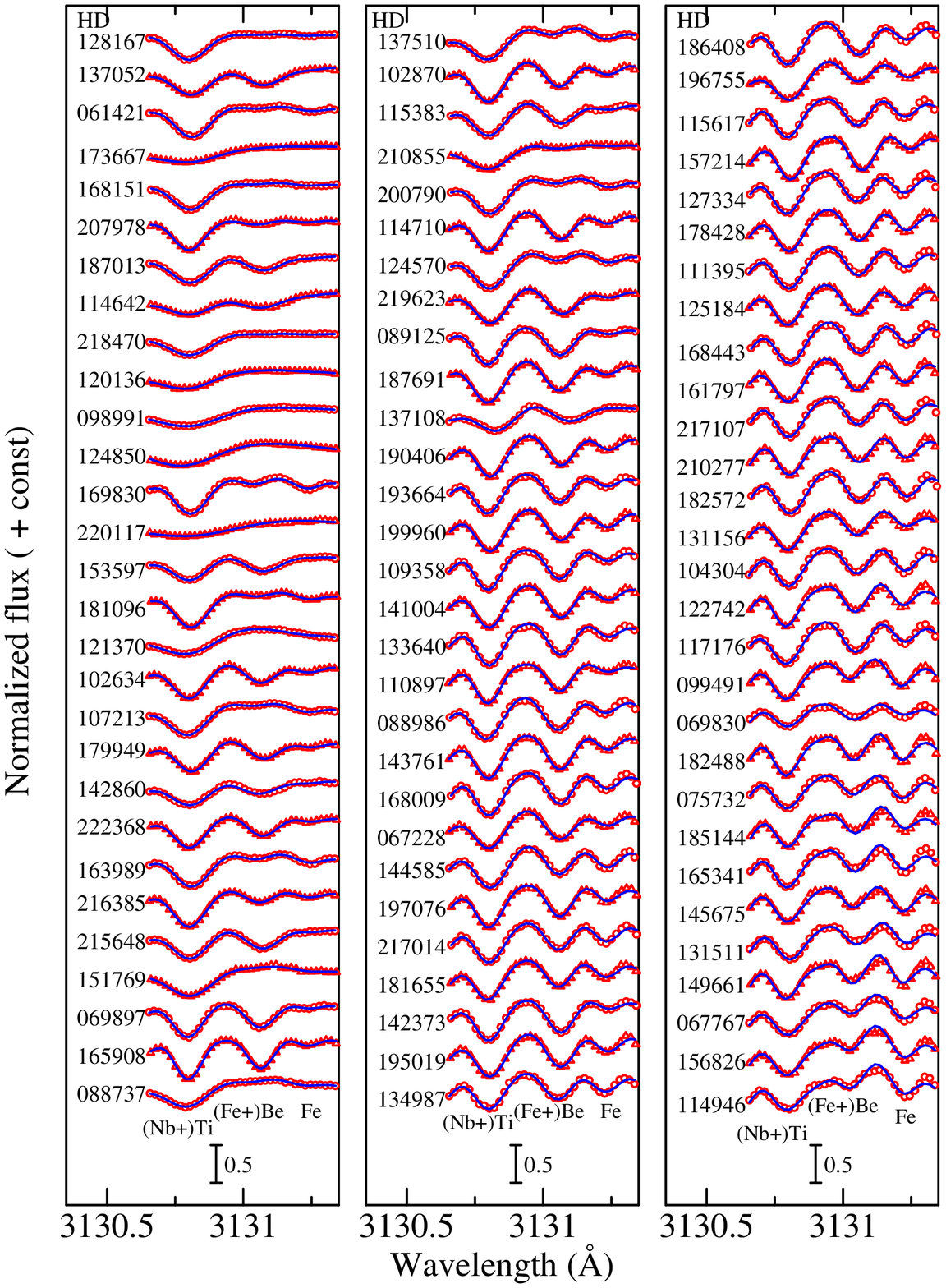}
  \end{center}
\caption{
Synthetic spectrum fitting in the 3130.65--3131.35~$\rm\AA$ region 
for 87 standard stars of F, G, and K type. Each of the spectra 
are arranged in the decreasing order of $T_{\rm eff}$. 
Indicated on the left to the spectrum is the HD number. 
Otherwise, the same as in figure 4.
}
\end{figure}

\setcounter{figure}{7}
\begin{figure}
  \begin{center}
    \FigureFile(120mm,60mm){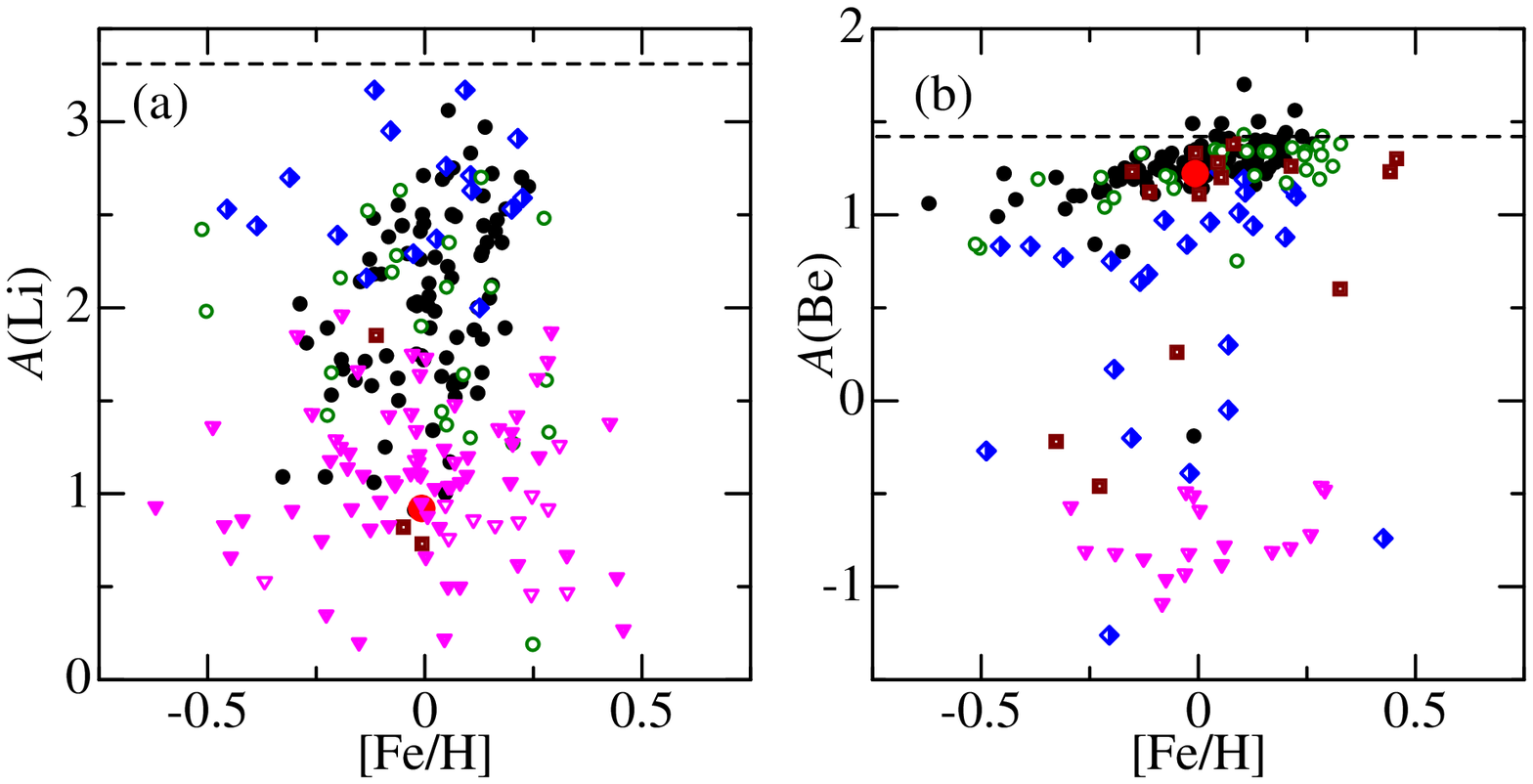}
  \end{center}
\caption{
Li and Be abundances plotted against the metallicity: 
(a) $A$(Li) vs. [Fe/H]; (b) $A$(Be) vs. [Fe/H]. 
The same meanings of the symbols as described in the caption of 
figure 2. Note that the (pink) downward triangles indicate
the upper limit values.
The meteoritic abundances (3.31 for Li, 1.42 for Be; cf. Anders \& Grevesse) 
are indicated by the horizontal dashed lines.

}
\end{figure}

\setcounter{figure}{9}
\begin{figure}
  \begin{center}
    \FigureFile(140mm,180mm){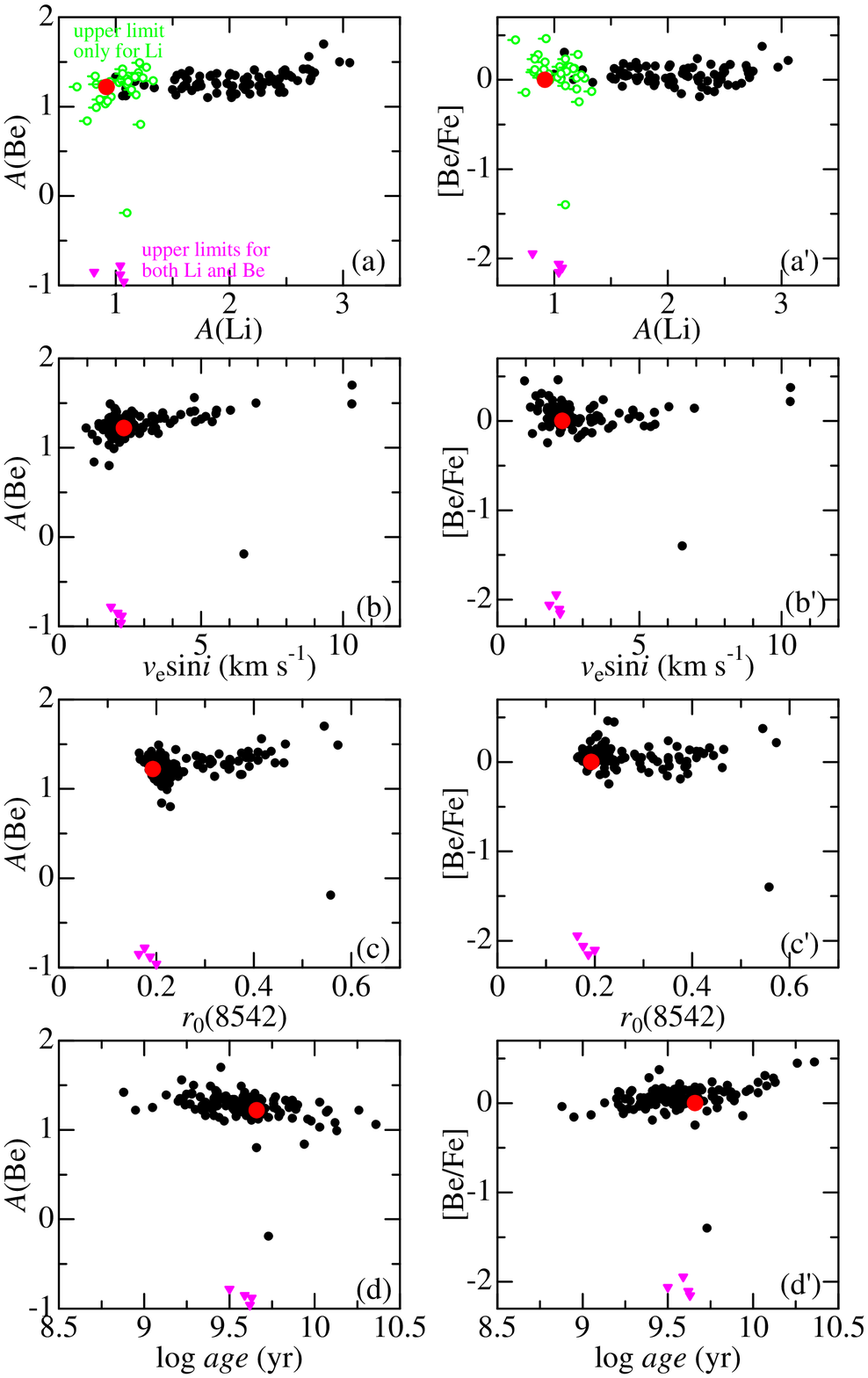}
  \end{center}
\caption{
Beryllium abundances of 118 solar analogs (and the Sun) plotted
against the Li abundance (from Paper I; panels a/a$'$), 
the projected rotational velocity (from Paper II; panels b/b$'$),
the residual flux at the Ca~{\sc ii} 8542 line center
(from Paper II; panels c/c$'$), and the age (from Paper I;
panels d/d$'$). The left panels are for $A$(Fe), while the 
right ones are for [Be/Fe]. Pink downward triangles 
indicate the upper limit values for both Be and Li abundances 
(four stars). Regarding the green open circles (with leftward 
ticks) in panels (a) and (a$'$), the upper limit values are relevant 
only for the Li abundances.
}
\end{figure}

\setcounter{table}{0}
\setlength{\tabcolsep}{3pt}
\begin{table}[h]
\scriptsize
\caption{Adopted data of atomic and molecular lines.}
\begin{center}
\begin{tabular}{cccrr  cccrr cccrr}\hline\hline
Species & $\lambda$ & $\chi$ & $\log gf$ & $\eta_{0}$ &  
Species & $\lambda$ & $\chi$ & $\log gf$ & $\eta_{0}$ &  
Species & $\lambda$ & $\chi$ & $\log gf$ & $\eta_{0}$ \\  
        & ($\rm\AA$)&  (eV)  &           &        &
        & ($\rm\AA$)&  (eV)  &           &        &
        & ($\rm\AA$)&  (eV)  &           &        \\
\hline
OH&3128.060 &0.54 &$-$2.73 &1.17 &Cr~{\sc i}&3129.774 &2.71 &$-$2.46 &0.04 &Tm~{\sc ii}&3131.257 &0.00 &+0.28&1.36 \\
OH&3128.101 &0.21 &$-$3.18 &0.79 &Fe~{\sc i}&3129.802 &2.88 &$-$2.80 &1.66 &OH&3131.329 &1.94 &$-$1.80 &0.59 \\
OH&3128.237 &0.44 &$-$3.32 &0.36 &Cr~{\sc i}&3129.858 &2.97 &$-$2.27 &0.04 &OH&3131.384 &1.68 &$-$3.22 &0.04 \\
Sc~{\sc ii}&3128.269 &3.46 &$-$0.17 &0.25 &Y~{\sc ii}&3129.935 &3.42 &+0.98&1.04 &OH&3131.393 &1.68 &$-$3.29 &0.03 \\
OH&3128.286 &0.21 &$-$2.29 &6.24 &OH&3129.938 &1.61 &$-$1.84 &1.05 &Fe~{\sc ii}&3131.395 &3.81 &$-$3.46 &0.89 \\
OH&3128.290 &0.44 &$-$3.14 &0.55 &Gd~{\sc ii}&3129.968 &1.17 &$-$0.20 &0.11 &OH&3131.423 &0.96 &$-$2.16 &1.84 \\
V~{\sc ii}&3128.305 &2.38 &$-$0.87 &1.83 &Zr~{\sc i}&3130.056 &0.52 &$-$0.95 &0.02 &Fe~{\sc i}&3131.455 &2.47 &$-$3.60 &0.60 \\
OH&3128.356 &1.72 &$-$2.61 &0.14 &OH&3130.125 &0.84 &$-$2.15 &2.40 &Fe~{\sc i}&3131.460 &2.47 &$-$4.53 &0.07 \\
OH&3128.376 &1.72 &$-$3.48 &0.02 &Ti~{\sc i}&3130.158 &1.98 &$-$0.46 &0.64 &OH&3131.501 &0.50 &$-$2.73 &1.25 \\
CH&3128.393 &0.56 &$-$1.34 &1.55 &Fe~{\sc i}&3130.205 &3.33 &$-$2.60 &1.05 &Ni~{\sc i}&3131.525 &3.19 &$-$2.90 &0.09 \\
CH&3128.393 &0.56 &$-$1.30 &1.69 &V~{\sc ii}&3130.258 &0.35 &$-$0.74 &145.76 &Cr~{\sc ii}&3131.533 &4.17 &$-$1.51 &2.95 \\
Dy~{\sc ii}&3128.406 &1.31 &+0.35&0.56 &OH&3130.280 &0.25 &$-$1.97 &11.80 &Cr~{\sc ii}&3131.548 &4.18 &$-$1.46 &3.25 \\
Ti~{\sc ii}&3128.490 &3.90 &$-$0.02 &4.88 &Ce~{\sc ii}&3130.340 &0.53 &$-$0.15 &0.49 &OH&3131.686 &1.74 &$-$2.51 &0.17 \\
OH&3128.518 &0.79 &$-$2.49 &1.23 &CH&3130.370 &0.03 &$-$1.96 &1.07 &Ni~{\sc i}&3131.702 &3.31 &$-$1.61 &1.42 \\
OH&3128.525 &0.10 &$-$3.30 &0.75 &Ti~{\sc i}&3130.376 &1.43 &$-$1.56 &0.15 &OH&3131.710 &1.74 &$-$3.50 &0.02 \\
Cr~{\sc ii}&3128.545 &4.76 &$-$2.07 &0.25 &OH&3130.407 &1.76 &$-$3.49 &0.02 &Fe~{\sc ii}&3131.724 &4.08 &$-$2.10 &11.85 \\
Gd~{\sc ii}&3128.569 &1.13 &$-$0.10 &0.15 &Be~{\sc ii}&3130.421 &0.00 &$-$0.17 &9.97 &OH&3131.755 &0.96 &$-$3.26 &0.15 \\
Ti~{\sc i}&3128.627 &2.10 &$-$0.49 &0.47 &OH&3130.432 &1.76 &$-$2.37 &0.23 &Co~{\sc i}&3131.826 &1.74 &$-$2.59 &0.10 \\
Mn~{\sc ii}&3128.640 &6.67 &$-$1.15 &0.03 &OH&3130.473 &1.61 &$-$3.22 &0.04 &Cr~{\sc ii}&3132.053 &2.49 &$-$0.32 &1335.65 \\
Cr~{\sc ii}&3128.692 &2.43 &$-$0.72 &589.75 &Fe~{\sc i}&3130.476 &3.58 &$-$2.96 &0.28 &Zr~{\sc i}&3132.063 &0.54 &+0.02&0.22 \\
Cu~{\sc i}&3128.692 &4.97 &$-$0.71 &0.05 &Mn~{\sc ii}&3130.551 &6.49 &$-$1.15 &0.04 &OH&3132.186 &0.90 &$-$2.32 &1.46 \\
V~{\sc ii}&3128.694 &2.37 &$-$0.67 &2.97 &Cr~{\sc ii}&3130.568 &5.33 &$-$2.46 &0.03 &Co~{\sc i}&3132.212 &0.10 &$-$2.52 &3.09 \\
Ni~{\sc i}&3128.728 &1.95 &$-$4.29 &0.04 &OH&3130.570 &0.68 &$-$1.78 &7.81 &CH&3132.281 &0.49 &$-$1.38 &1.61 \\
Y~{\sc ii}&3128.737 &3.38 &+0.81&0.76 &Cr~{\sc i}&3130.585 &3.56 &$-$1.97 &0.02 &CH&3132.281 &0.49 &$-$1.33 &1.82 \\
OH&3128.783 &0.90 &$-$2.29 &1.55 &Mn~{\sc i}&3130.637 &4.27 &$-$1.01 &0.11 &Mn~{\sc i}&3132.289 &4.33 &$-$0.50 &0.32 \\
Fe~{\sc i}&3128.897 &1.56 &$-$3.05 &13.14 &CH&3130.648 &0.03 &$-$1.55 &2.72 &Mn~{\sc i}&3132.405 &3.38 &$-$2.09 &0.06 \\
OH&3128.975 &1.94 &$-$1.77 &0.63 &Nb~{\sc ii}&3130.780 &0.44 &+0.41&5.85 &Er~{\sc ii}&3132.517 &1.40 &+0.51&0.63 \\
Co~{\sc i}&3129.006 &0.51 &$-$2.93 &0.53 &Ti~{\sc ii}&3130.810 &0.01 &$-$1.53 &370.74 &Fe~{\sc i}&3132.519 &3.21 &$-$1.10 &42.62 \\
Fe~{\sc ii}&3129.009 &3.97 &$-$2.90 &2.39 &Gd~{\sc ii}&3130.813 &1.16 &$-$0.08 &0.15 &OH&3132.584 &1.61 &$-$2.00 &0.72 \\
Ti~{\sc i}&3129.070 &2.12 &+0.49&4.31 &Ce~{\sc ii}&3130.871 &1.09 &+0.48&0.69 &Ce~{\sc ii}&3132.592 &0.30 &$-$0.54 &0.32 \\
OH&3129.094 &0.90 &$-$3.24 &0.17 &CH&3130.928 &0.00 &$-$3.14 &0.07 &Mo~{\sc i}&3132.595 &0.00 &$-$0.13 &0.82 \\
Zr~{\sc ii}&3129.153 &0.53 &$-$0.12 &20.87 &OH&3130.928 &1.91 &$-$2.52 &0.12 &V~{\sc ii}&3132.596 &2.90 &$-$1.07 &0.41 \\
Fe~{\sc i}&3129.182 &2.45 &$-$4.46 &0.09 &OH&3130.933 &0.68 &$-$3.13 &0.34 &Fe~{\sc i}&3132.649 &3.24 &$-$1.60 &12.77 \\
Cr~{\sc i}&3129.210 &3.56 &$-$1.11 &0.17 &Mn~{\sc ii}&3131.015 &6.11 &$-$1.22 &0.07 &Ti~{\sc i}&3132.710 &2.00 &$-$0.77 &0.30 \\
Ni~{\sc i}&3129.300 &0.28 &$-$2.75 &44.23 &Mn~{\sc i}&3131.037 &3.77 &$-$1.73 &0.06 &Mn~{\sc i}&3132.789 &4.27 &$-$0.50 &0.36 \\
Fe~{\sc i}&3129.334 &1.48 &$-$2.49 &55.00 &Fe~{\sc i}&3131.043 &2.85 &$-$2.52 &3.34 &V~{\sc ii}&3132.810 &2.51 &$-$1.53 &0.31 \\
Co~{\sc i}&3129.482 &1.88 &$-$2.02 &0.28 &Be~{\sc ii}&3131.066 &0.00 &$-$0.47 &5.00 &Cr~{\sc i}&3132.822 &3.12 &$-$0.19 &3.42 \\
OH&3129.539 &0.52 &$-$2.62 &1.57 &Th~{\sc ii}&3131.070 &0.00 &$-$1.56 &0.01 &OH&3132.845 &1.95 &$-$2.46 &0.13 \\
Ti~{\sc i}&3129.636 &1.44 &$-$1.45 &0.19 &Zr~{\sc i}&3131.109 &0.52 &$-$0.40 &0.09 &Ni~{\sc ii}&3132.865 &2.87 &$-$3.65 &0.79 \\
Nb~{\sc ii}&3129.652 &1.32 &$-$0.94 &0.04 &Os~{\sc i}&3131.116 &1.84 &+0.05&0.10 &OH&3132.866 &0.69 &$-$3.24 &0.27 \\
Mn~{\sc i}&3129.667 &3.77 &$-$1.51 &0.10 &Cr~{\sc i}&3131.212 &3.11 &$-$0.66 &1.18 &&&&&\\
Zr~{\sc ii}&3129.764 &0.04 &$-$0.65 &16.37 &Fe~{\sc i}&3131.235 &2.18 &$-$3.30 &2.14 &&&&&\\
\hline
\end{tabular}
\end{center}
Notes:\\ 
Atomic and molecular line data we used for spectrum synthesis.
Five kinds of data are presented for each line:
(1) species designation, (2) line wavelength, (3) lower excitation potential, 
(4) logarithm of lower statistical weight times oscillator strength,
and (5) line(center)-to-continuum opacity ratio at $\tau_{5000} = 0.2$
computed for the solar case (good indicator of the observed line-strength).
The data of (1)--(4) are based on the line list of Primas et al. (1997; cf.
table 2 therein)
\end{table}

\setcounter{table}{1}
\setlength{\tabcolsep}{3pt}
\begin{table}[h]
\small
\caption{Be abundances of 18 planet-host stars.} 
\begin{center}
\begin{tabular}
{cccccccl}\hline \hline
Star& $T_{\rm eff}$& $\log g$ & [Fe/H] & $A$(Be)& [Be/Fe]& $A$(Li)&Remark\\
     & (K) & (cm~s$^{-2}$) & (dex) & (dex) & (dex) & (dex) & \\
\hline
HIP~26381&5518 &4.47 &$-$0.45 &1.22 &+0.45 &($<0.66$)&solar analog\\
HIP~53721&5819 &4.19 &$-$0.02 &1.23 &+0.03 &1.75 &solar analog\\
HIP~59610&5829 &4.34 &$-$0.06 &1.23 &+0.07 &1.62 &solar analog\\
HIP~90004&5607 &4.42 &$-$0.02 &1.33 &+0.13 &($<1.17$)&solar analog\\
HIP~96901&5742 &4.32 &+0.08 &1.37 &+0.07 &($<1.06$)&solar analog\\
\hline
HD~120136&6420 &4.21 &+0.28 &($<-0.46$)&($<-1.96$)&($<1.71$)&standard ($T_{\rm eff}>6000$)\\
HD~169830&6355 &4.14 &+0.17 &($<-0.81$)&($<-2.20$)&($<1.35$)&standard ($T_{\rm eff}>6000$)\\
HD~179949&6294 &4.49 &+0.23 &1.10 &$-$0.35 &2.59 &standard ($T_{\rm eff}>6000$)\\
HD~143761&5832 &4.25 &$-$0.22 &1.20 &+0.20 &1.42 &standard ($6000>T_{\rm eff}>5500$)\\
HD~217014&5779 &4.30 &+0.20 &1.17 &$-$0.25 &1.27 &standard ($6000>T_{\rm eff}>5500$)\\
HD~195019&5768 &4.11 &+0.04 &1.35 &+0.09 &1.44 &standard ($6000>T_{\rm eff}>5500$)\\
HD~134987&5766 &4.37 &+0.28 &1.32 &$-$0.18 &($<0.92$)&standard ($6000>T_{\rm eff}>5500$)\\
HD~168443&5593 &4.16 &+0.06 &1.34 &+0.06 &($<0.76$)&standard ($6000>T_{\rm eff}>5500$)\\
HD~217107&5575 &4.18 &+0.31 &1.26 &$-$0.27 &($<1.26$)&standard ($6000>T_{\rm eff}>5500$)\\
HD~210277&5567 &4.44 &+0.25 &1.32 &$-$0.15 &($<0.46$)&standard ($6000>T_{\rm eff}>5500$)\\
HD~117176&5466 &3.80 &$-$0.11 &1.12 &+0.01 &1.85 &standard ($5500>T_{\rm eff}$)\\
HD~75732&5328 &4.58 &+0.46 &1.30 &$-$0.38 &($<0.27$)&standard ($5500>T_{\rm eff}$)\\
HD~145675&5309 &4.45 &+0.44 &1.23 &$-$0.43 &($<0.55$)&standard ($5500>T_{\rm eff}$)\\
\hline
\end{tabular}
\end{center}
Notes:\\ 
The data of atmospheric parameters and Li abundances were taken from
Paper I (for solar analog stars) as well as Takeda et al. (2005) and 
Takeda and Kawanomoto (2005) (for standard stars).
See electronic tables E1 and E2 for more detailed information concerning
these stars as well as for the complete data for the whole 205 stars.
\end{table}

\setcounter{table}{2}
\setlength{\tabcolsep}{3pt}
\begin{table}[h]
\small
\caption{Non-LTE effect on Be abundance determination
from the Be~{\sc ii} 3131.066 line.} 
\begin{center}
\begin{tabular}
{cccccccccrr}\hline \hline
Model & $T_{\rm eff}$ & $\log g$ & [Fe/H] & $v_{\rm t}$ & $A_{\rm input}$ & ($EW^{\rm L}$)&
$EW^{\rm N}_{0}$ & $EW^{\rm N}_{-3}$ & $\Delta^{\rm N}_{0}$ & $\Delta^{\rm N}_{-3}$ \\
 & (K) & (cm~s$^{-2}$) & (dex) & (km~s$^{-1}$) & (dex) &(m$\rm\AA$) & (m$\rm\AA$) & 
(m$\rm\AA$) & (dex) & (dex) \\
\hline
T50G40P00 & 5000& 4.0& 0.00& 1.5& 1.15& (62.5) & 62.2 & 60.5 & +0.01& +0.03\\
T55G40P00 & 5500& 4.0& 0.00& 1.5& 1.15& (80.0) & 80.0 & 78.9 &  0.00& +0.01\\
T60G40P00 & 6000& 4.0& 0.00& 1.5& 1.15& (89.7) & 88.3 & 87.7 & +0.02& +0.03\\
T65G40P00 & 6500& 4.0& 0.00& 1.5& 1.15& (85.5) & 79.4 & 79.1 & +0.09& +0.09\\
\hline
\end{tabular}
\end{center}
Notes:\\ 
The first five columns give the atmospheric parameters of each model.
$A_{\rm input}$ in column 6 is the input (solar) beryllium abundance 
(Anders \& Grevesse 1989), with which three kinds of equivalent widths 
[LTE, non-LTE ($\log k = 0$), and non-LTE ($\log k = -3$); $k$ is the
reduction factor of classical H~{\sc i} collision rates] are computed
as presented in column 7--9. Column 10 and 11 give the non-LTE corrections
for each case of $k$. $\Delta^{\rm N}_{0}$ is defined as 
$A^{\rm N}_{0} - A^{\rm L}_{0}$, where $A^{\rm N}_{0}$ and $A^{\rm L}_{0}$ 
are the non-LTE ($k=1$) and LTE abundances inversely derived from 
$EW^{\rm N}_{0}$, respectively. (Note that $A^{\rm N}_{0}$ is essentially 
equal to $A_{\rm input}$.)  
Regarding $\Delta^{\rm N}_{-3}$, a similar definition applies.
\end{table}

\end{document}